\documentclass[12pt,preprint]{aastex}

\def\gs{\mathrel{\raise0.35ex\hbox{$\scriptstyle >$}\kern-0.6em
\lower0.40ex\hbox{{$\scriptstyle \sim$}}}}
\def\ls{\mathrel{\raise0.35ex\hbox{$\scriptstyle <$}\kern-0.6em
\lower0.40ex\hbox{{$\scriptstyle \sim$}}}}

\shortauthors{Owen et al}
\shorttitle{Survey of A2125 III.}

\begin{document}

\title{A Deep Radio Survey of A2125 III.}
\title{The Cluster Core -- Merging and Stripping}

\author{
Frazer\,N.\ Owen,\altaffilmark{1,2}, 
W.\,C.\ Keel,\altaffilmark{3,2},
Q.\, D.\ Wang\altaffilmark{4},
M.\,J.\ Ledlow\altaffilmark{5,2,7}, G.\,E.\ Morrison,\altaffilmark{6}
}
\altaffiltext{1}{National Radio Astronomy Observatory, P.\ O.\ Box O,
Socorro, NM 87801 USA. The National Radio Astronomy Observatory is
facility of the National Science Foundation operated under cooperative
agreement by Associated Universities Inc.}
\altaffiltext{2}{Visiting astronomer, Kitt Peak National Observatory,
National Optical Astronomy Observatories, operated by AURA, Inc.,
under cooperative agreement with the National Science Foundation.}
\altaffiltext{3}{Dept. of Physics \& Astronomy, University of Alabama,
Tuscaloosa, AL 35487 USA}
\altaffiltext{4}{Department of Astronomy, University of Massachusetts,
710 North Pleasant Street, MA 01003}
\altaffiltext{5}{Gemini Observatory, Southern Operations Center, AURA,
Casilla 603, La Serena, Chile}
\altaffiltext{6}{IfA, University of Hawaii, and Canada-France-Hawaii
  Telescope, Kamuela, HI 96743}
\altaffiltext{7}{Deceased 5 June 2004. We shall miss his cheerfulness,
unfailing good sense, and scientific industry.}

\setcounter{footnote}{8}

\begin{abstract}
We use radio, near-IR, optical, and X-ray observations to examine
dynamic processes in the central region of the rich galaxy cluster
Abell 2125. In addition to the central triple of E and  cD galaxies, including
members of both major dynamical subsystems identified from a redshift
survey, this region features a galaxy showing strong evidence for
ongoing gas stripping during a high-velocity passage through the gas
in the cluster core. The disk galaxy C153 exhibits a plume stretching toward
the cluster center seen in
soft X-rays by {\it Chandra}, parts of which are 
also seen in [O II] emission and
near-UV continuum light. {\it HST} imaging shows a distorted disk, with
star-forming knots asymmetrically distributed and remnant spiral
structure possibly defined by dust lanes. The stars and ionized gas in its disk
are kinematically decoupled, demonstrating that pressure stripping
(possibly turbulent as well as ram) must be important, and that tidal 
disruption
is not the only mechanism at work. Comparison of the gas properties 
seen in the X-ray and optical data on the plume highlight significant
and poorly-known features of the history of stripped gas in the intracluster
medium which could be clarified through further observations of this system.
The nucleus of C153 also hosts an AGN, shown by the
weak and distorted extended radio emission and a radio compact
core. The unusual strength
of the stripping signatures in this instance is likely related to the very
high relative
velocity of the galaxy with respect to the intracluster medium, during
a cluster/cluster
merger, and its passage very near the core of the cluster. An additional
sign of recent dynamical events is the diffuse starlight asymmetrically
placed about the central triple in a cD envelope. Transient and
extreme dynamical events 
as seen in Abell 2125 may be important drivers of galaxy evolution in the
cores of rich clusters.

\end{abstract}


\section{Introduction}

Progressively richer data in the optical and X-ray have revealed
that clusters of galaxies, in addition to being important sites
for galaxy evolution, are themselves evolving systems. This
is seen in the evidence for cluster mergers, as shown by
galaxy position/velocity arrays and substructure in the
hot intracluster medium (ICM). Particularly intriguing has been 
evidence that such events on a cluster-wide scale might affect the
individual galaxies, as manifested via star formation, occurrence
of nuclear activity, and/or modifications due to the effects of tidal 
and ram-pressure forces. We have conducted an intensive study of Abell 2125,
a cluster of richness class 4 at $z=0.247$, originally motivated by
fractions of blue galaxies and radio detections which are exceptional for
its redshift and richness. As discussed by \citet{m03} and
\citet{o03b}
(Paper II), A2125
appears to be a major merger in progress
seen close to the line-of-sight. Spatially the two
largest galaxy concentrations overlap in projection. Modeling suggests
a position angle to the line-of-sight of about 30$^\circ$ and that
the two systems are within 0.2 Myr of core passage. The projected scale
of the full A2125 concentration is about 5 Mpc, making the total extent
about 10 Mpc, consistent with a major merger. Most of the
radio-detected galaxies 
are distributed  uniformly in projection throughout this region. We
argued in Paper II that the radio emission from most of these
systems is due to star-formation activity and that these galaxies live
in intermediate density environments, more conducive to mergers,
interactions, and thus star-formation. The on-going major cluster
merger may also be enhancing the galaxy-galaxy interactions in 
group-like environments we are seeing. 

Additional aspects of the cluster merger appear to be active in the
richest parts of A2125. The core region has four,
fairly luminous radio sources ($> 10^{23}$ W Hz$^{-1}$). All the
evidence which we will present here suggests that, unlike most
rich clusters as we currently observe them, a complex interaction
is taking place which is the result of the merger of two dense 
sub-clumps in the A2125 system. Most striking, one of these radio
sources arises from a smaller galaxy in the process of losing
much of its interstellar medium to the ram pressure of a rapid
passage through the densest part of the ICM.
We  discuss here the optical, radio,
and X-ray evidence for this interpretation.

\section{Observations}

	The radio, optical and near-IR imaging are described in
more detail in previous papers. These
observations include 1) a deep VLA A+B configuration survey
\citep{o03a} (Paper I), 
2) deep wide-field imaging in U,B,V,R,I,J,H,K, and two narrow bands
at 8010 \AA\ and 9170 \AA\ using the KPNO 2.1m and 4m telescopes
\citep{o03a},
3 ) optical
spectroscopy using the KPNO 4m and WIYN telescopes \citep{o99,m03}, 
4) HST imaging in F606W and F814W  \citep{o03b} and 5) a {\it Chandra} 82ksec 
integration \citep{w03}.

	Besides these observations, we describe here further
VLA radio observations and new GMOS
spectroscopy from the GEMINI-N telescope as described below.

\subsection{VLA Observations}

Besides the 1400 MHz observations, the cluster core was observed
in the VLA D configuration on August 28, 2000 at 4860, 8460 and 14940
MHz to determine the radio spectrum of C153. The observations were
taken in the standard 50 MHz bandwidth continuum mode for total
integration times of 53, 39, and 31 minutes respectively for the
three frequencies. Calibrations were performed in the standard way
and imaged and cleaned with IMAGR, all in the AIPS package. 

\subsection{GEMINI GMOS Spectroscopy}

Spectra of C153 were obtained during two Gemini-N programs
(GN-2002A-Q-15 and GN-2003A-Q-14), using GMOS-N with 1" wide slits.
The first observation, using a 5" slit length, was on 12 March 2002
(observation N20020312S0015), giving one 
3000-second exposure. The slit was tilted by about 20$^\circ$ with
respect to the default direction (that is, perpendicular to the dispersion)
to follow the galaxy's projected
major axis. The grating R400-G5300 (400 lines per mm, 0.67 A/pixel) was
set with a center wavelength near 6300 \AA\ , giving coverage from [O II]
to the H$\alpha$ region. The small slit length arose from adding this
object to a program of identification of faint radio-selected objects.

The later spectrum used a 19" slit length, centered asymmetrically on C153 to 
accommodate primary targets. This logistical limitation means that the slit
did not encompass the plume seen in [O II] emission to the NW of the
galaxy disk, a feature that had not been found when the observations
were set up. This slit was likewise tilted
with respect to the CCD and dispersion, to sample along the major axis.
Two 2700-second exposures were done (observation N20030528S0100 and -101)
on 28 May 2003. These was centered near 5200 \AA\ , using
grating B600-G5303 (600 lines/mm, 0.45 A/pixel) so we cover from short-ward
of [O II] (to below 3200 \AA\ in the emitted frame) and red-ward slightly past 
[O III]. Atmospheric dispersion
compensation was not in use for either set of observations 
(air masses ranged from 1.4-1.6).

Reduction included flat fielding,
wavelength mapping using associated CuAr exposures, 
rebinning of the data to linear and orthogonal spatial and wavelength
scales, merging of data from the 3 CCDs, and sky subtraction.
Since no flux standard star was observed with the same setups during
these sessions, we calibrated the C153 spectrum to at least a relative
flux scale by reference to a Kitt Peak
4m observation (Miller et al. 2004) taken with a wide (2") slit. 

\section{Results}

\subsection{Radio/Optical Imaging}

\begin{figure}
\plottwo{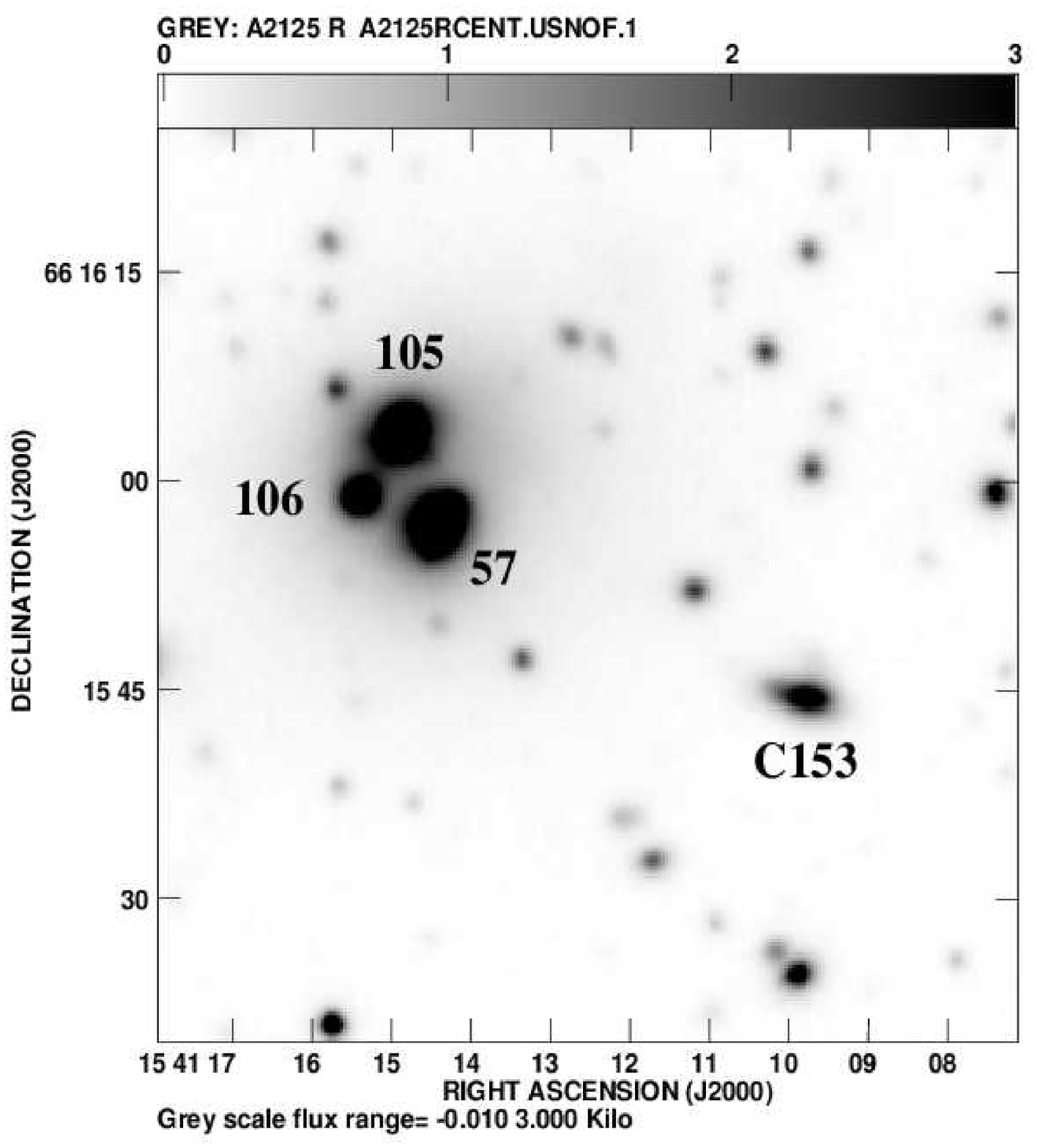}{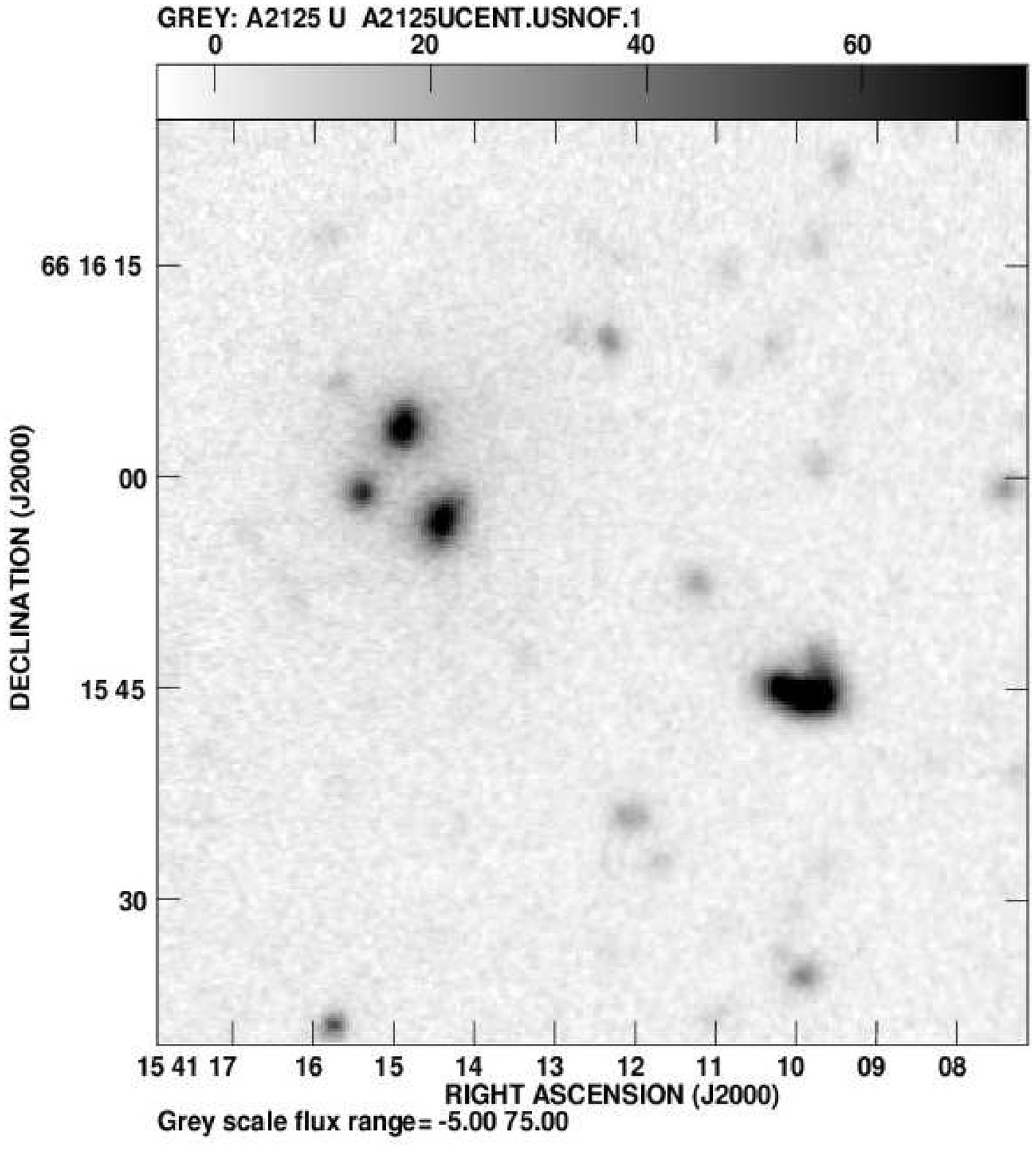}
\caption{R and U images of the cluster center. The left and right
panels contain the R and U images respectively. Note the
dramatic color change between C153 and the 
central galaxies (00057, 00105 and 00106). \label{URfigs}}
\end{figure}

\begin{figure}
\plotone{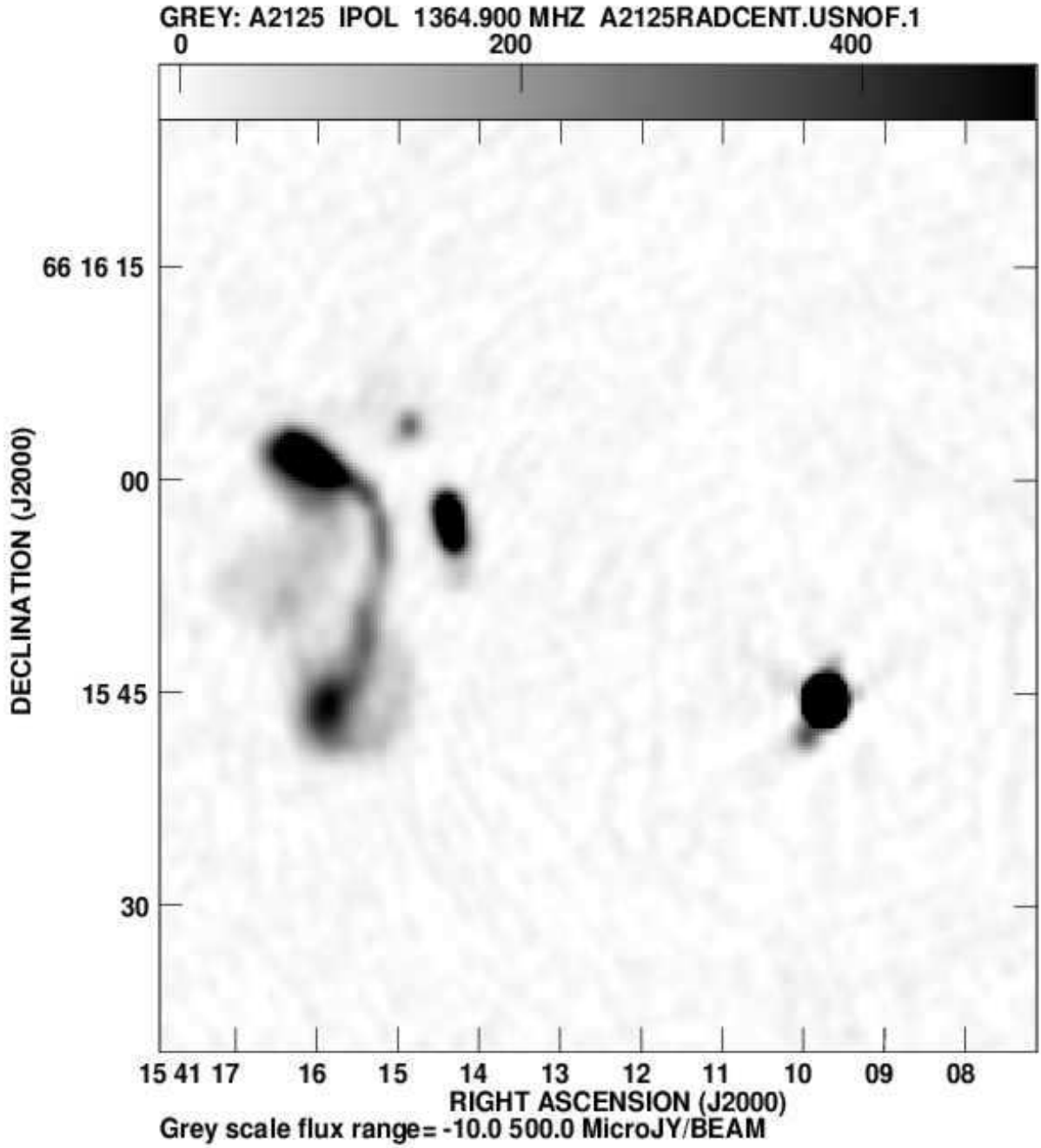}
\caption{The radio image covering the same region as the
R and U optical images \label{radio}}
\end{figure}

\begin{figure}
\plotone{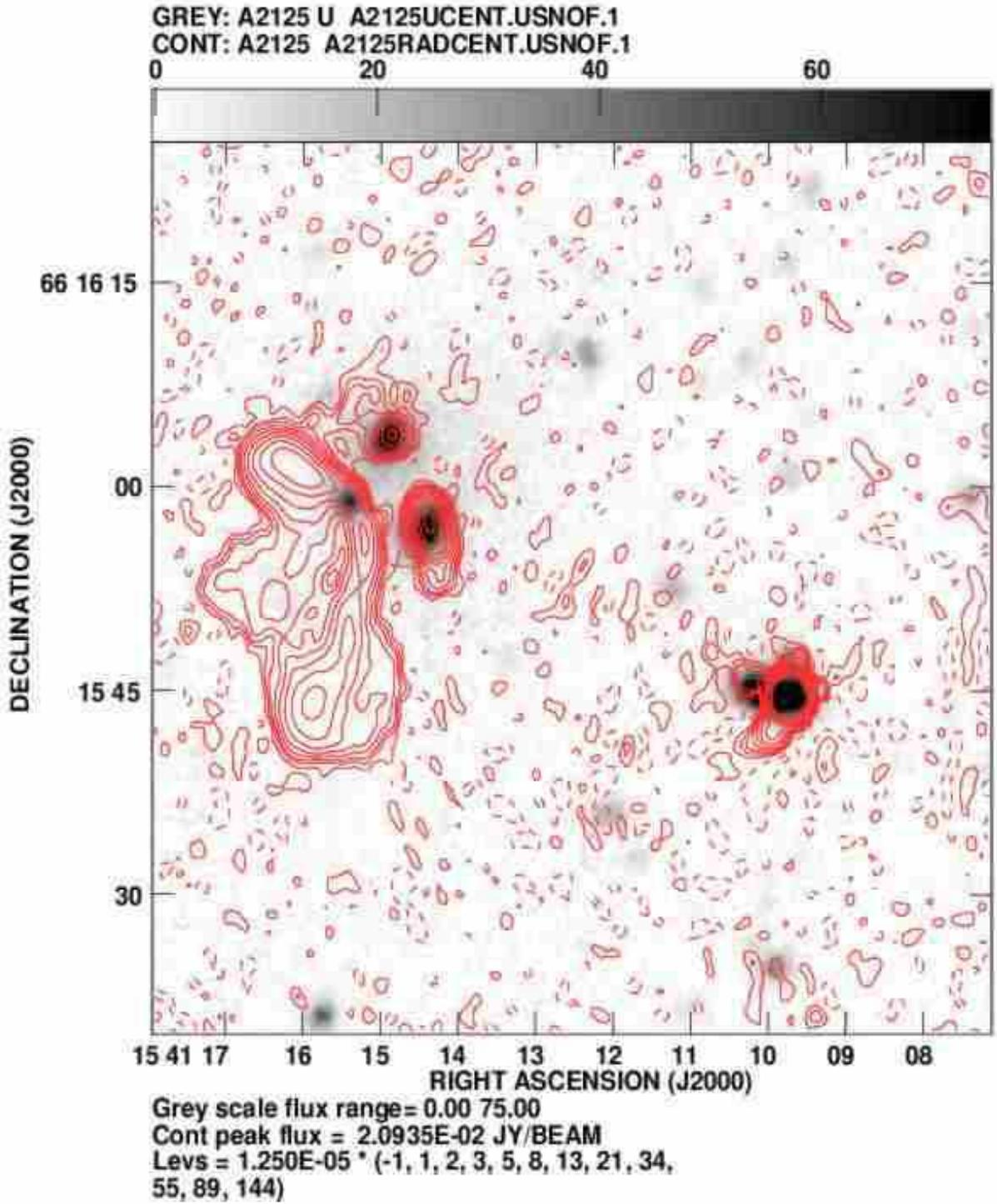}
\caption{The radio image, plotted as red contours, overlaid
on the optical U image. All four of the four central galaxies
have extended radio structure. \label{RO}}
\end{figure}

	In figure~\ref{URfigs}, we show the R and U images of the
cluster core from the KPNO 4m MOSAIC imaging. The central triple
system of the cluster is near (15 41 15, 66 16 00). Also note the 
galaxy C153 (source 00047 from Paper II), near (15 41 10, 66 15 45);
this galaxy will be the subject of much of the discussion in this
paper. The central triple of cD galaxies dominates the R image. 
In contrast, C153 is the brightest object in the field at U.
In figure~\ref{radio},
we show the radio image of the same field. Each of the three central
galaxies has an extended radio source. However, C153 has the brightest
peak, in addition to some extended structure of its own. In figure~\ref{RO},
we show the radio image as contours overlaid on the optical U
image. Although C153 is dominated by a point source, note the radio
extension along both the galaxy's major and minor axes. Also note the
extended emission associated with each of the three central galaxies
and the bent tails associated with the SE galaxy, object
00106. 

In table 1, we list some basic properties of the four central
galaxies from Paper II, using Gunn-Oke apertures for the optical
magnitudes. In columns 1 and 2, we give the source number and
coordinate name; in columns 3 and 4, the radio size and luminosity
in kpc and log W Hz$^{-1}$, the absolute magnitude in a Gunn-Oke
aperture (radius=13.1 kpc) is given in column 7; Columns 8, 9, 10 give
the result of the SED fitting including the spectral type
($8=$ continuous star formation, $1=$ star-burst), the characteristic
time since the beginning of the event and the Calzetti
extinction. Column 11 gives the 
redshift (e implies an emission line redshift). Please see paper II
for the details. The spectroscopic data for this quadruple
system reveals that
two of the central galaxies (00105 and 00106) have radial velocities
near the peak in the main cluster distribution. However, 00057 (the
SW member of the triple) and C153 both have radial velocities about
1400 km/s higher than the other two galaxies, in the velocity
range dominated by the less populous secondary cluster component of Paper II. 
The bent radio source
associated with 00106 also is consistent with this system moving at high speed
with respect to the surrounding medium.

\begin{figure}
\plotone{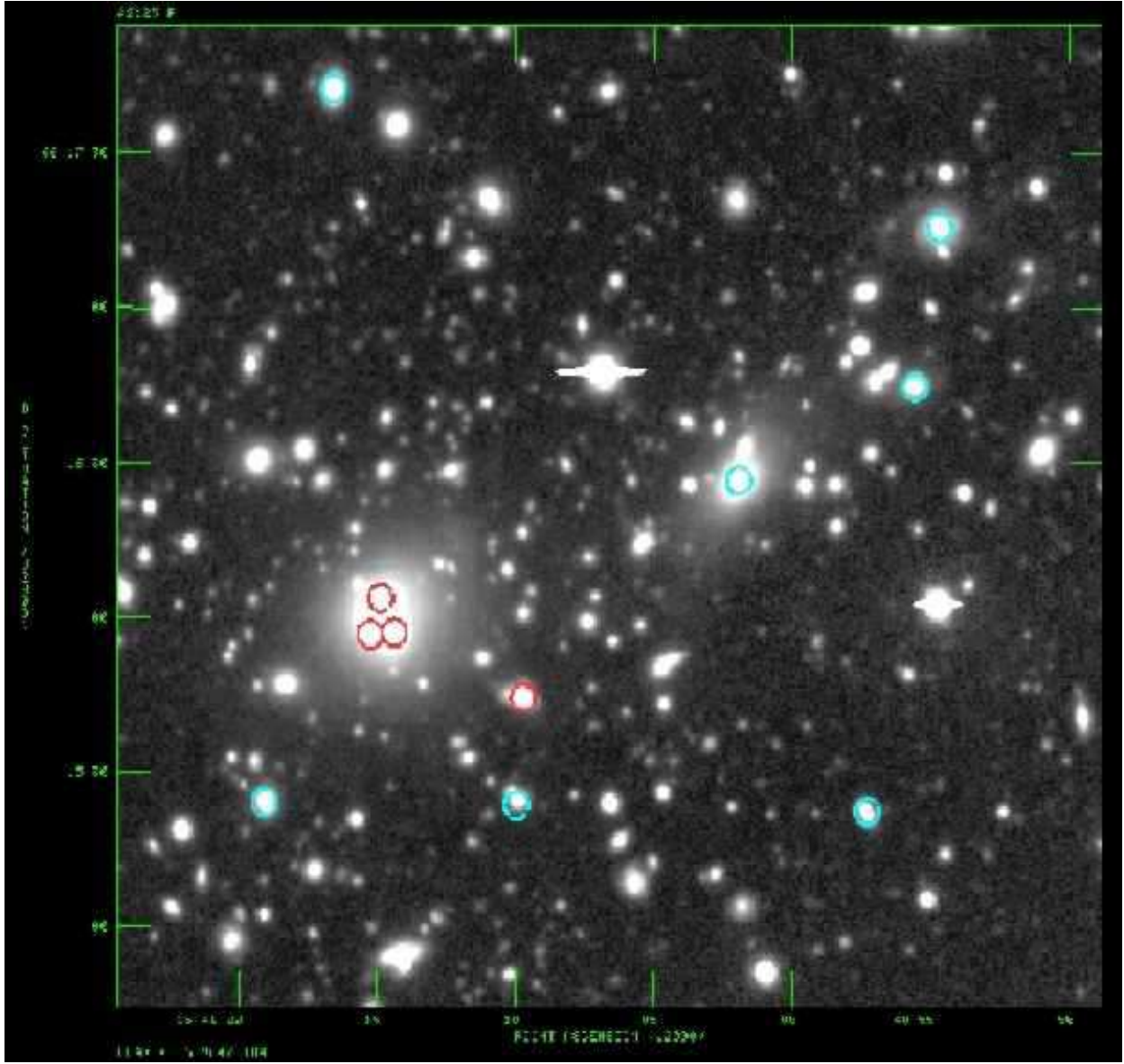}
\caption{The R-band image of the cluster core shown displayed
with a normal stretch to show all the individual galaxies. The
red circles shows the four four central galaxies discussed above.
The blue circles show lower radio luminosity systems also in the
field.  \label{H1}}
\end{figure}

\begin{figure}
\plotone{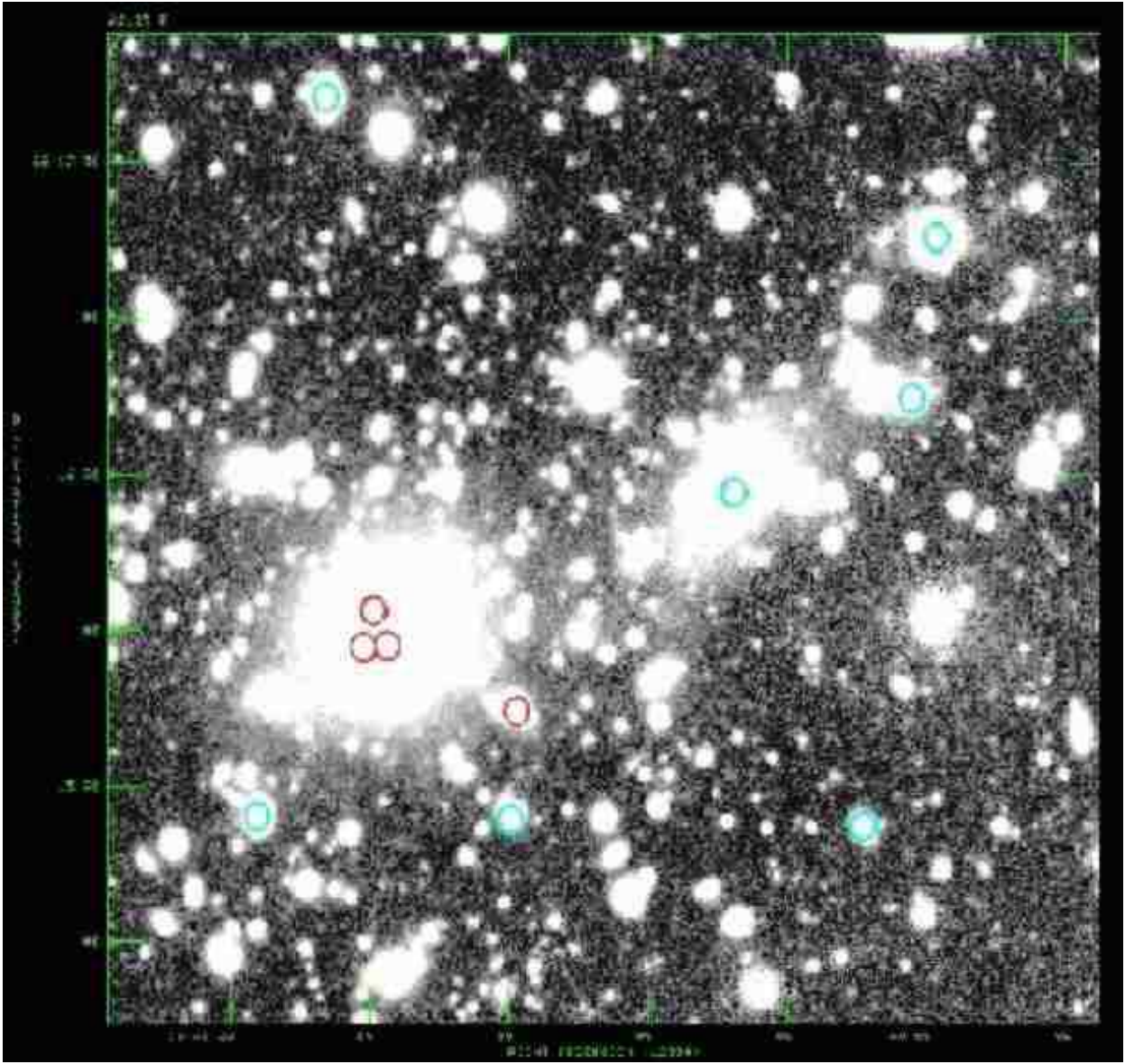}
\caption{The R-band image of the cluster core displayed to
show the more diffuse light in the cluster core. The circles
are the same as in figure~\ref{H1} \label{H2}}
\end{figure}

	On a slightly larger scale the cluster  shows evidence
of complex interactions at low light levels. In figures~\ref{H1}
and \ref{H2}, we show the R MOSAIC image of the cluster core with
two different stretches to show how the lower surface brightness
emission relates to the galaxies. Note that the three central
galaxies do not lie at the center of the diffuse light. Also note
the large halo around the galaxy about two arcmin north and west
of the central triple. This galaxy is radio galaxy 00039 and {\it Chandra}
X-ray source X065. The diffuse X-ray emission of the main
cluster also is elongated along a line connecting the central triple
and 00039 \citep{w03}. Once again on this larger scale there is
evidence of a complex interaction between multiple components of the
cluster and a situation not likely to come from a relaxed equilibrium.

As a sanity check on the reality of this diffuse light, given the complex
overlapping envelopes of the three central galaxies, we used iterative
application of the STSDAS {\it ellipse} and {\it bmodel} tasks to derive
symmetric models for the individual galaxies. This procedure converged
to a set of profiles for the galaxies which indicate that the diffuse
light must arise from something else - either very strong isophotal
distortions of the galaxies or a photometrically distinct component
of starlight. The X-ray brightness profile also shows this excess to the
northwest \citep{w03}.

The global stellar populations determined from the SEDs also are of
interest, tracing
broad features of the galaxies' histories. As listed in Paper II, the
broadband SED of C153
fits a continuous star-forming model with an age of 3.5 Gyr and a
Calzetti extinction law with an $A_V=1.0$. This system was the only
one we found in the cluster that fits this type of model
SED. Repeating the 
fits using smaller aperture sizes, C153 was always best fit
by a continuous star-formation model but with somewhat higher
extinction and shorter durations ($< 1$ Gyr). We will examine its star-forming
history in more detail below, using GMOS spectroscopy. 
The other three galaxies 
all were fit best by single burst models with ages between 5.5 and 9.5
Gyr. Such models basically look like simple old stellar populations,
noting that the WMAP consensus cosmology had an age of 10.5 Gyr at $z=0.25$.
The optical spectra of these three central objects show no emission lines.  

\begin{figure}
\plottwo{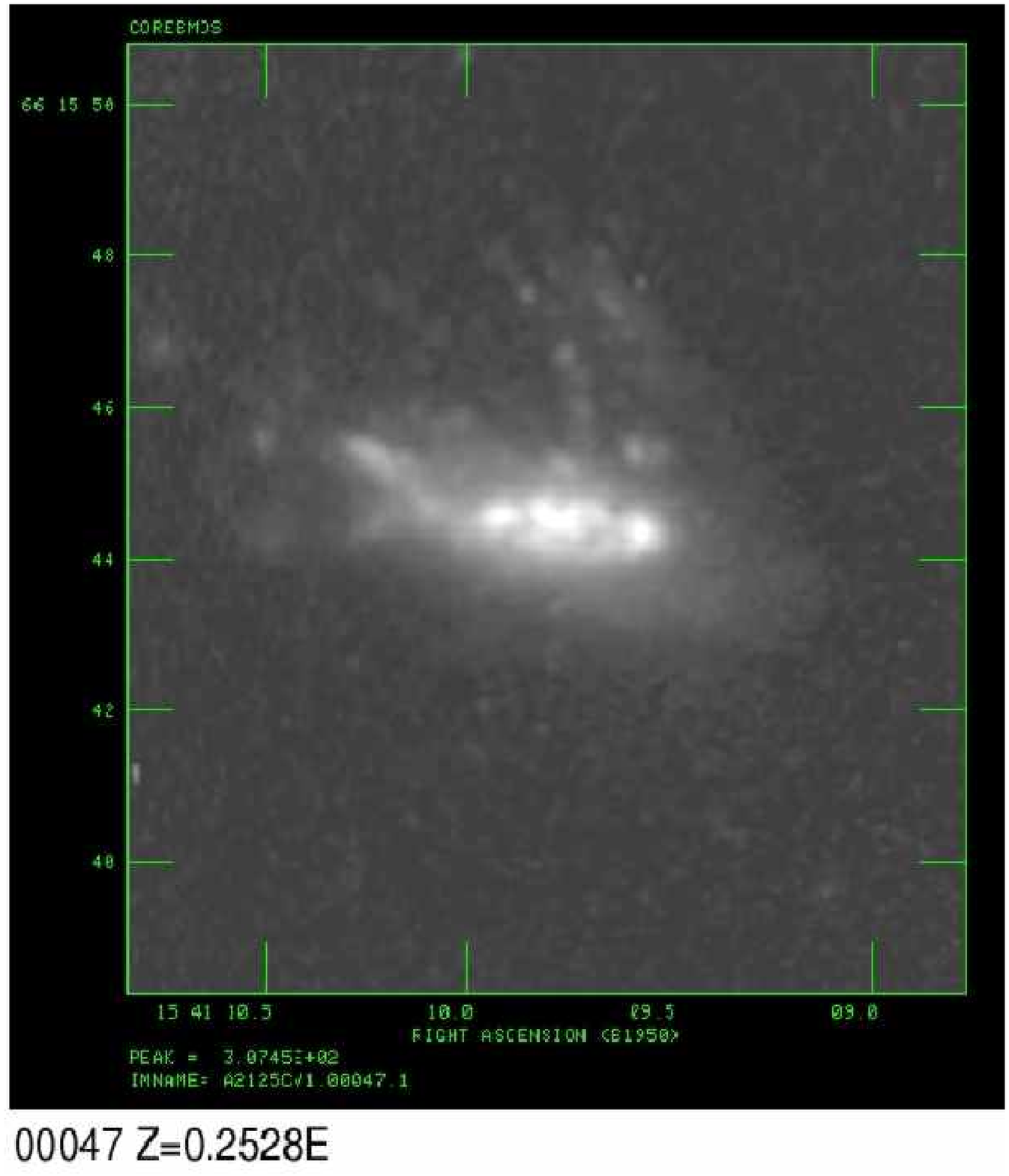}{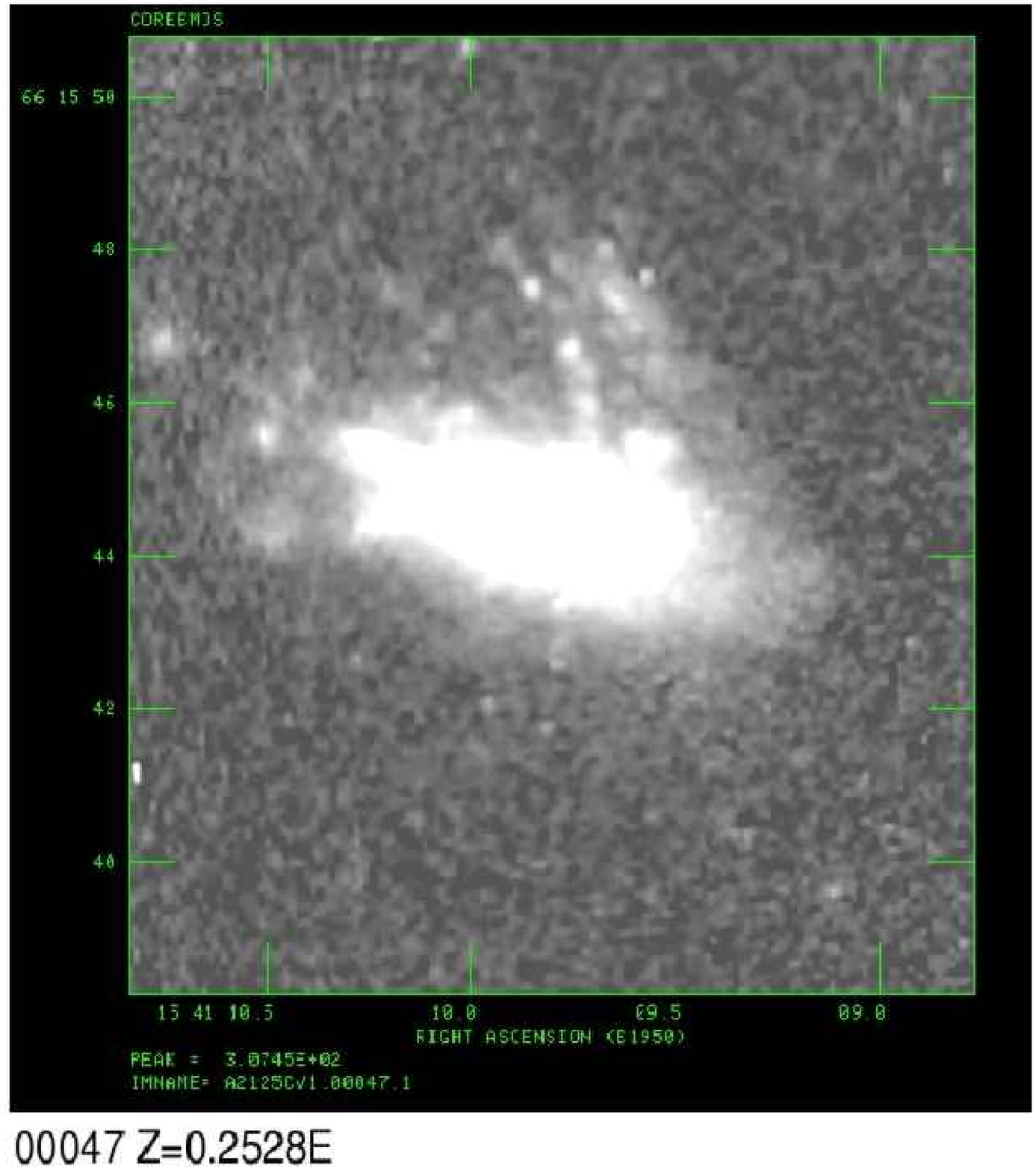}
\caption{The HST V Image of C153 displayed with two different
transfer functions \label{V47}}
\end{figure}

\begin{figure}
\plotone{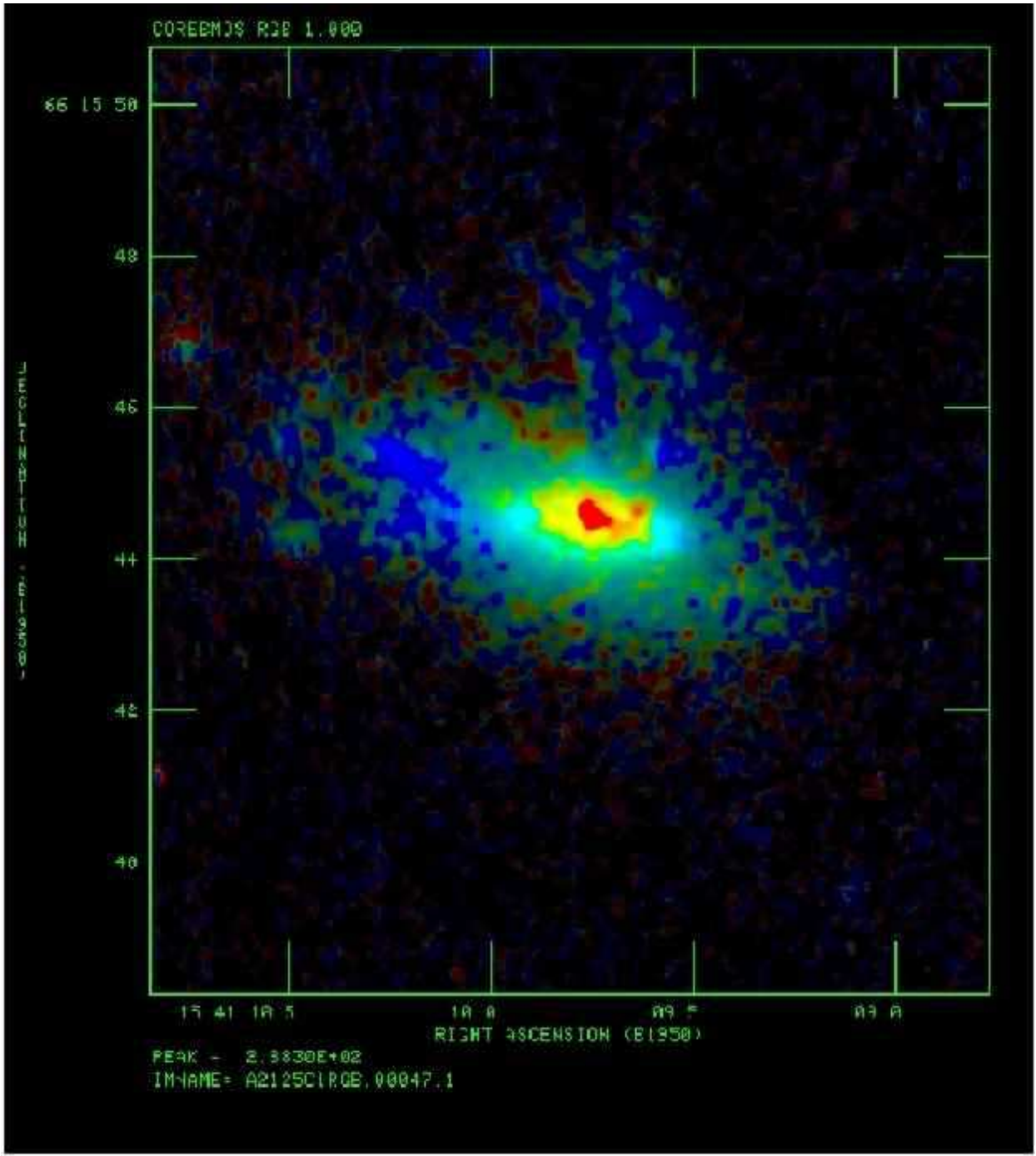}
\caption{A ``true color'' HST image of C153 made from the HST F606W
and F814W images. Note the redder colors in the galaxy nucleus which
suggest extinction from dust. \label{V47COL}}
\end{figure}

\begin{figure}
\plotone{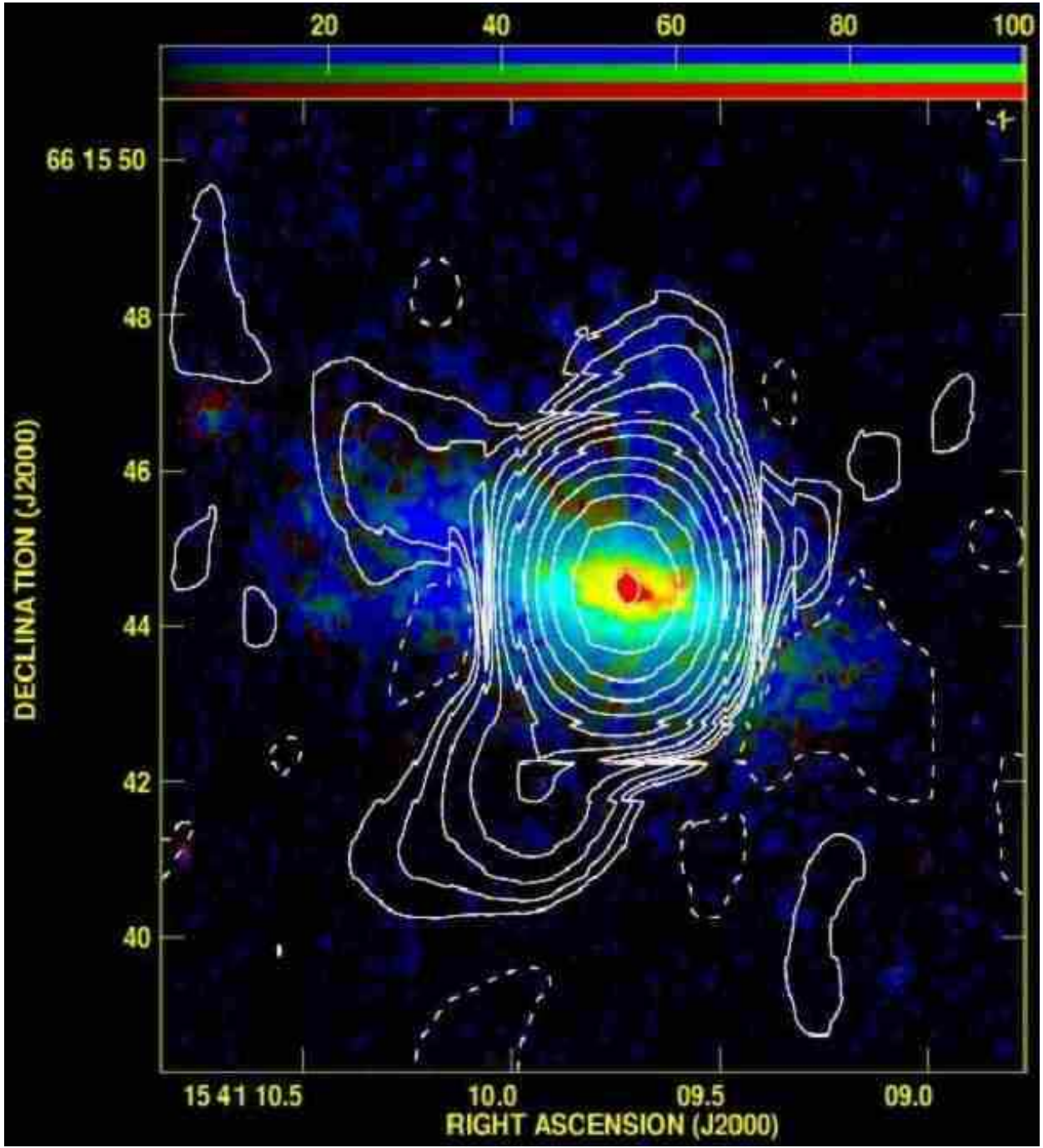}
\caption{True color HST image of A2125 with radio contours
overlaid \label{VRAD47COL}}
\end{figure}

\begin{figure}
\plotone{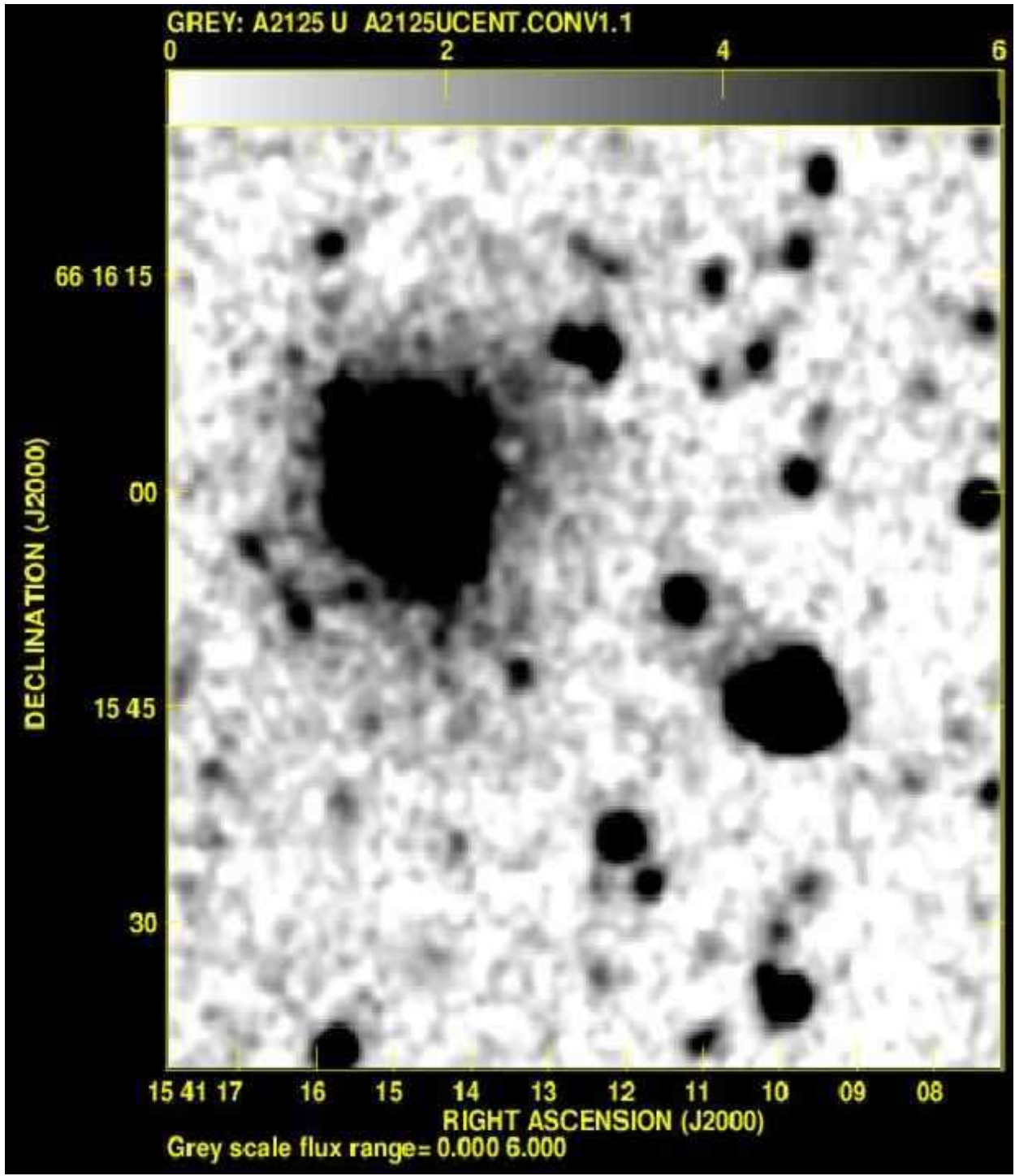}
\caption{The U-band image of the region around the cluster
core.\label{CU}}
\end{figure}

\begin{figure}
\plotone{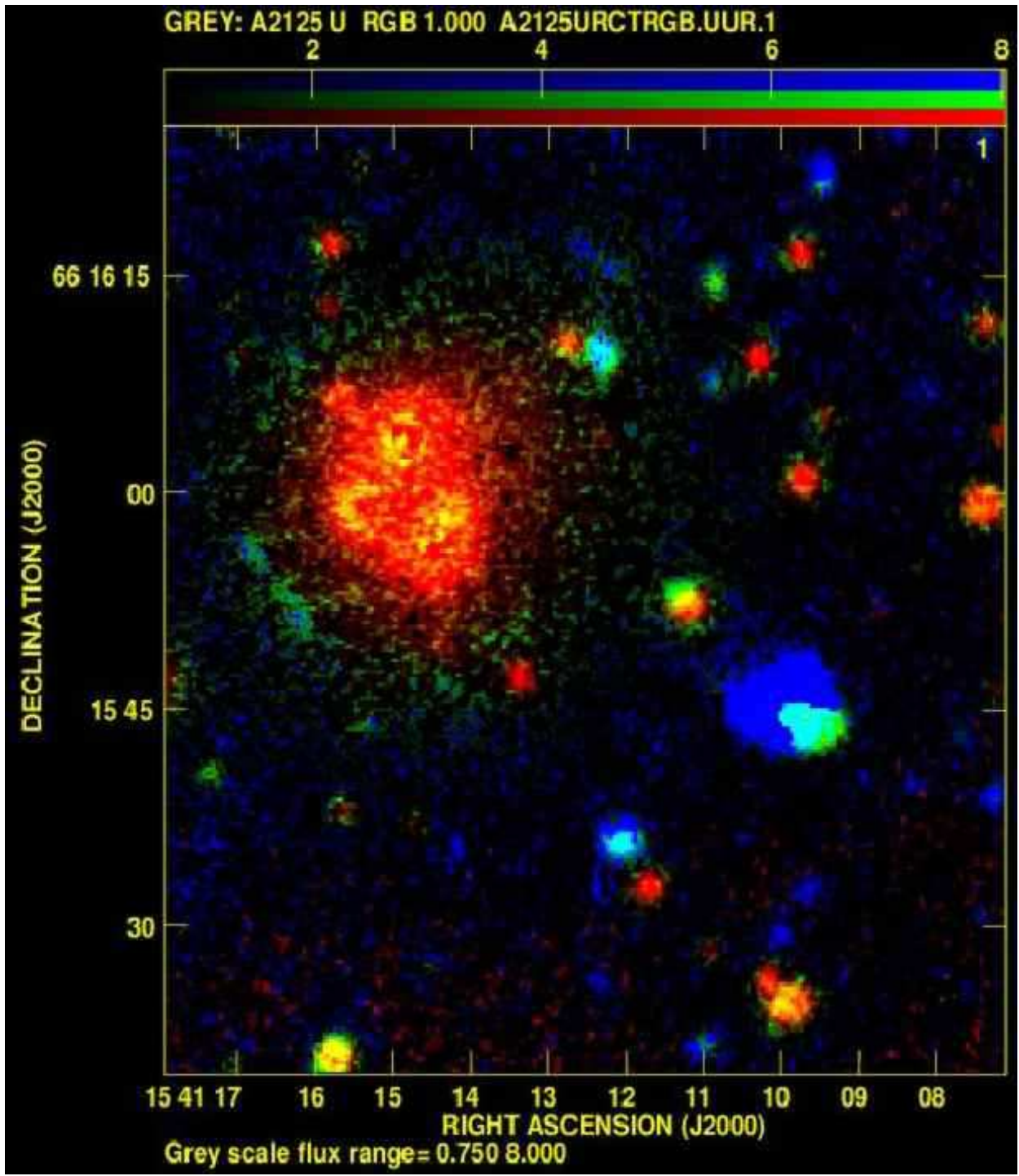}
\caption{A U and R ``true-color'' image of the core of the
cluster. \label{CCOL}}
\end{figure}

	From the ground, C153  appears to be a late type galaxy of some
type. However, in figure~\ref{V47} we show the WFPC2 image of the
system. Clearly this system is not a normal galaxy.  In
figure~\ref{V47COL}, we show a ``true color'' image of the system made
from the two HST images. The colors seem likely to be primarily due
to continuum emission. The observed equivalent widths of emission
lines in the core of the galaxy account for 3.2\% and 9.8\% of light in the
F606W and F814W filters respectively. Away from the nucleus where we
do not have spectroscopy the percentages could be more significant. 
The central core of the galaxy is red with respect to the outer more
diffuse regions, suggesting dust-obscuration.  Thus the galaxy appears to
be the combination of a
heavily dust-obscured nuclear region along with less obscured blue
light from young stars coming from the complex outer region of the
system. In figure~\ref{VRAD47COL}, we show the 1.5 arcsec resolution
radio image overlaid on the HST color image.  The extended radio
emission seems generally aligned with the blue light, except to the
south of the nucleus, while the
unresolved radio core is coincident with the red nuclear region.
	In figure~\ref{CU}, we show the U-band image of the
region containing the four central objects stretched to show the
lower surface brightness emission. One can see a faint broad
region of emission extending about 10 arcsec from C153 back
toward the central triple. One can also see two approximately
linear features near (15 41 16.5, 66 15 53) and (15 41 12.5,
66 16 17) which more or less tangential with respect to the central
diffuse emission. In figure~\ref{CCOL}, we show a ``true
color'' image made from the U and R images of the same region.
One can see that these two features are quite blue. It seems
possible that these two features are gravitational arcs, especially
since the presence of arcs in such low-redshift clusters is
correlated with the kinds of substructure expected during mergers
\citep{b95,a98,t04}.

\begin{figure}
\plotone{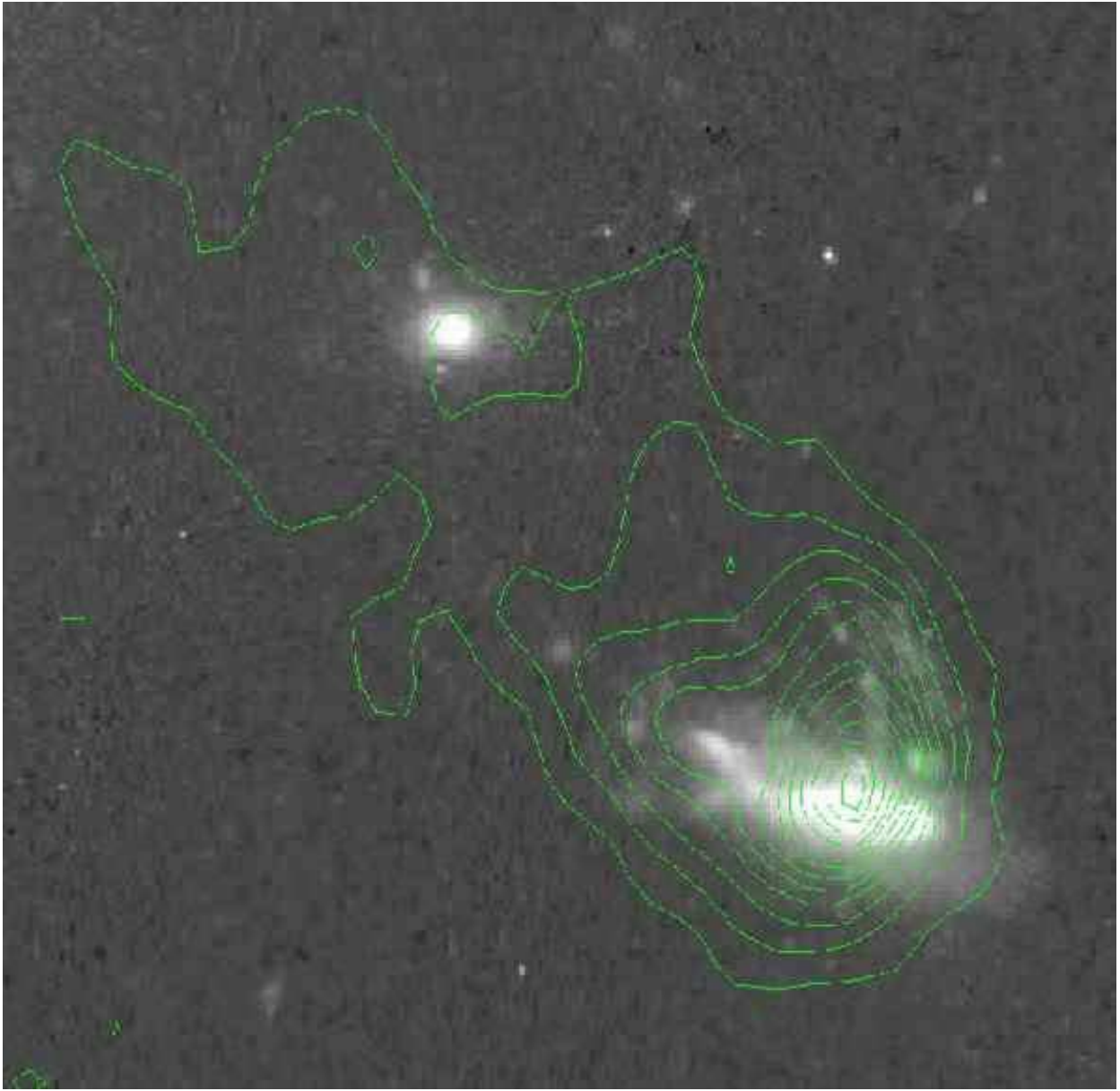}
\caption{The HST V image of C153 with the KPNO MOSAIC [OII]
image overlaid as contours. \label{OII}}
\end{figure}

	As part of our MOSAIC imaging program we found that
there was a narrow-band filter (center wavelength 4653 \AA, 
FWHM 52\AA) in the KPNO set which was very
well matched to [OII] at the redshift of A2125. In figure~\ref{OII},
we show the HST WFPC2 image with contours of the continuum subtracted
[OII] image overlaid. Note the the similarity of the U band continuum
image in figure~\ref{CU} and the [OII] image. Also note the northern
extension of the [OII] emission aligns with the filamentary structure seen
on the HST image. Overall, we see a similar structure in each of the
$U$, [O II], and soft X-ray bands - a plume or tail stretching to the 
northwest from C153, extending at least 80 kpc in projection from the
[O II] image,
and perhaps stretching, further based on the X-ray image (e.g. 
figure~ref{X02}) to become
confused with weaker sources in the
cluster. This feature furnishes strong evidence that
many of the oddities of C153 are related to an intense episode of
ISM stripping.
	
\subsection{{\it Chandra} Imaging}

\begin{figure}
\plotone{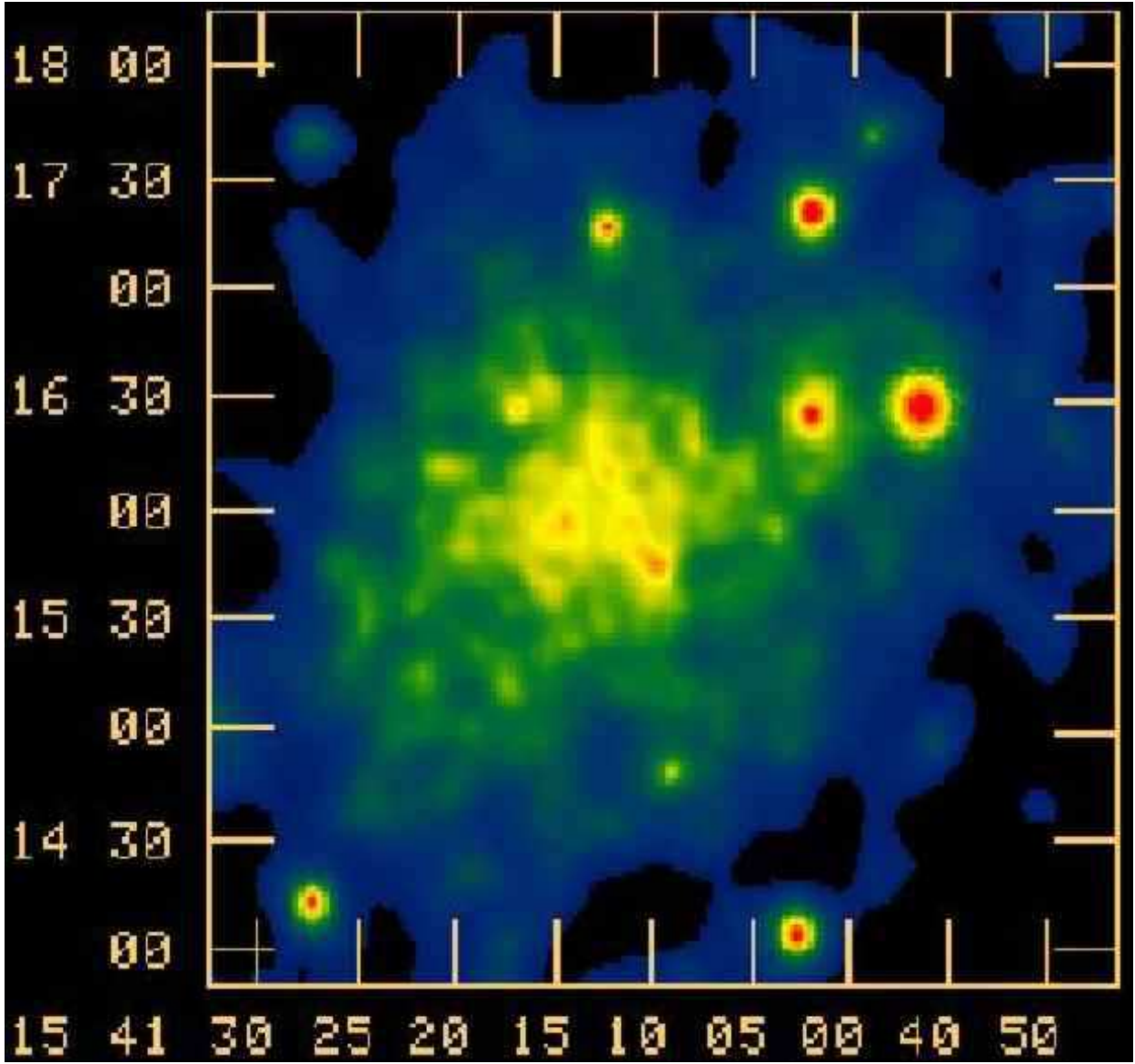}
\caption{The cluster core of A2125 imaged by {\it Chandra}. This
field is larger than the previous images. C153 is located 
near (15 41 09, 66 15 44).\label{X}}
\end{figure}

\begin{figure}
\plotone{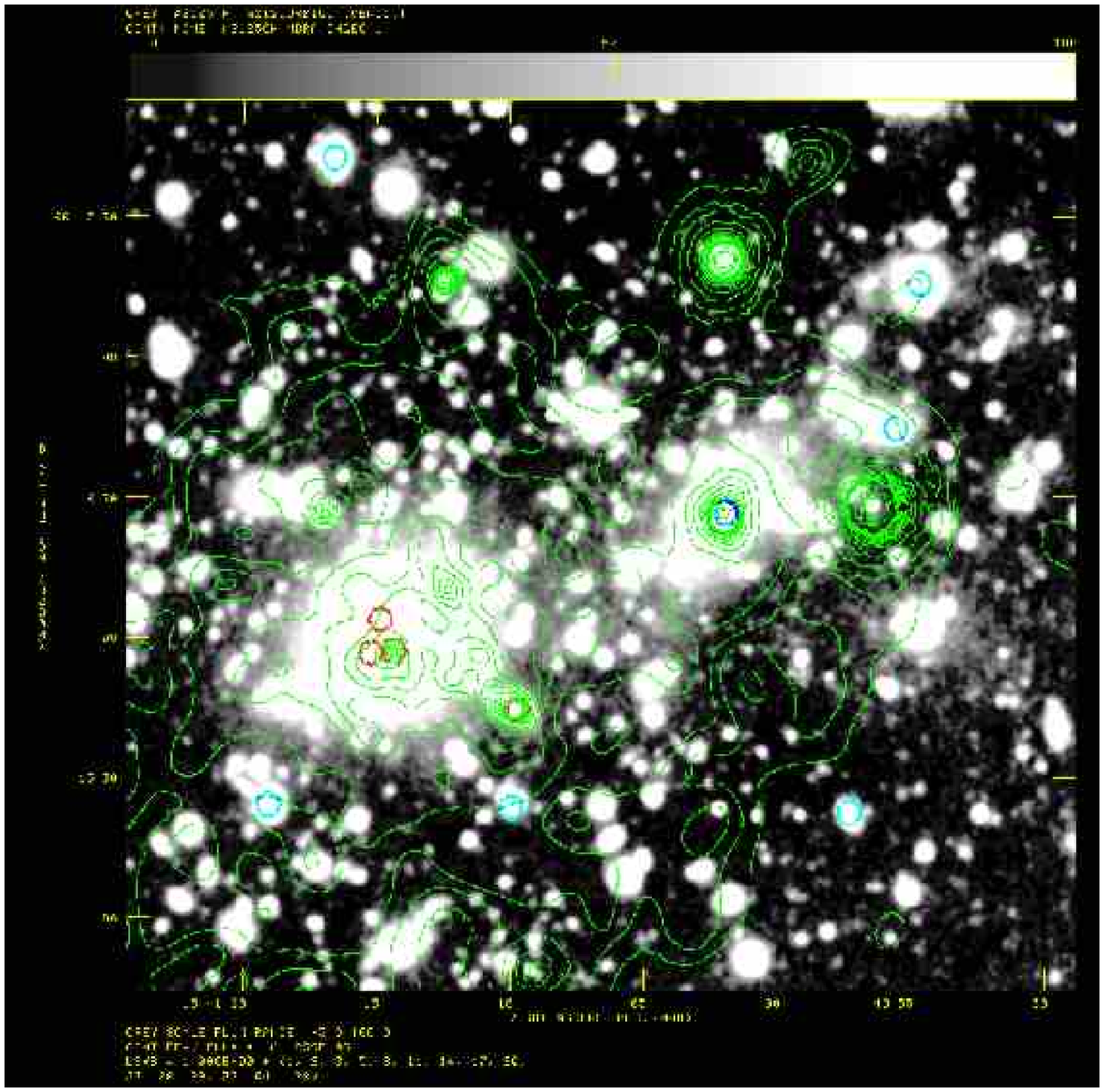}
\caption{The R image of the cluster core with the {\it Chandra}
image overlaid as contours \label{XO}}
\end{figure}

\begin{figure}
\plottwo{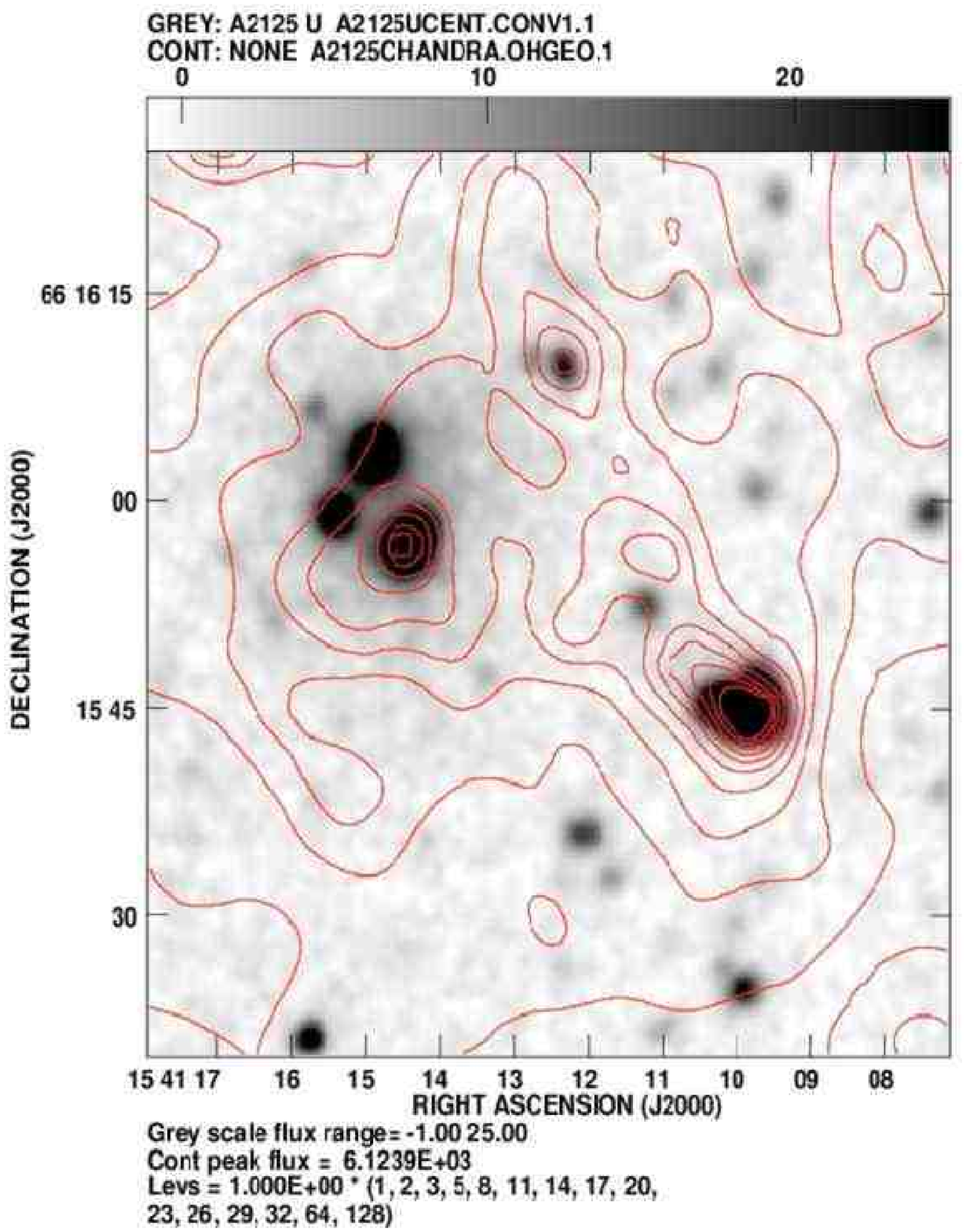}{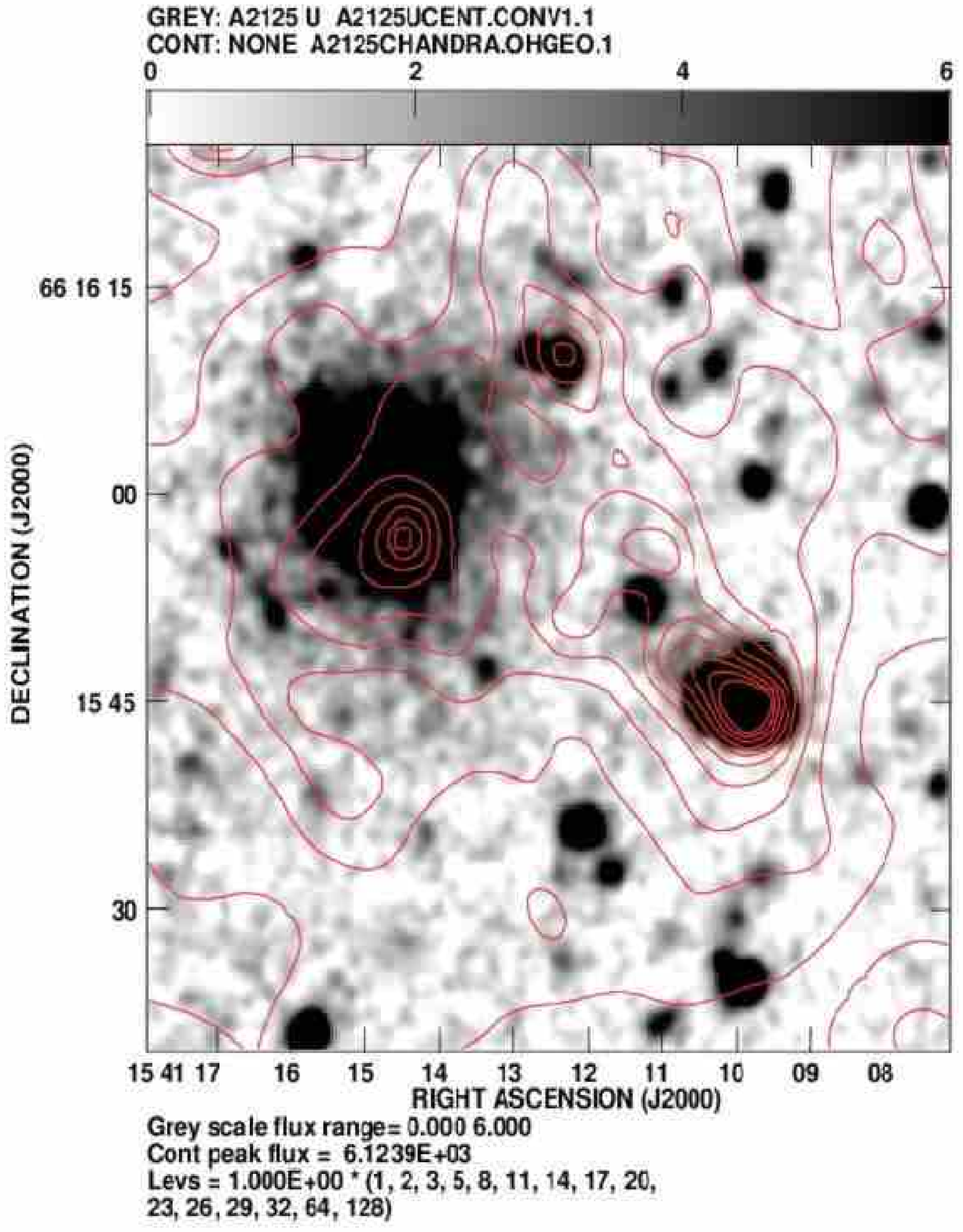}
\caption{The U image of the central part of the cluster core using 
two different transfer functions with the {\it Chandra} image overlaid. 
as contours. Note the correspondence of the diffuse near-UV light
with the inner part of the trail or plume of X-ray emission from C153.
 \label{XO2}}
\end{figure}

	As described in \citet{w03}, A2125 was observed for 82 ksec
by {\it Chandra} in 2001. In figure~\ref{X}, we present the 
0.5 to 8 kev {\it Chandra} image
of the core region. In this image the X-ray core appears both
complex and unusual. In figure~\ref{XO} we display the
wider field, R-band image of the cluster core with the X-ray contours
overlaid. ne can see the overall association of the X-ray emission
with the lower surface brightness R-band emission. In figure~\ref{XO2} we
show the U image of the cluster core, with two different transfer
functions, with the {\it Chandra} image overlaid. From the left hand
panel, one can see the general
alignment of the X-ray tail of C153 with the diffuse U-band emission
(and thus the [OII] emission). The length of the X-ray tail appears to
exceed the optical emission. However, in the right hand panel, one
can see a faint galaxy near (15 41 11.5,
66 16 14) appears also to coincide with a local peak in the emission.
Thus the length of the X-ray tail is at least 
$\sim 16$ arcsec or about 80 kpc
in projection. From the left hand panel of figure~\ref{XO2} one can
also see that the diffuse
emission at the core of the cluster is centered on the SW member of
the central triple. The two other core galaxies do not obviously have
associated X-ray emission.

\subsection{C153 GMOS Spectrum}

We examined clues to the stellar population in C153, using the spatially
integrated GMOS spectrum (covering a region 81 pixels long, or 5.8", by 1"
extended along the major axis). We considered primarily the
details of the spectral fit between emitted wavelengths 3200-4200 \AA\ ,
not only because of the range of diagnostic stellar lines in this
region, but to reduce sensitivity to reddening (which changes the
weighting of stellar populations) within the spectral range fitted.

\subsubsection{Stellar Population}

The GMOS spectrum includes numerous stellar absorption lines, especially
the high-order Balmer features. Their relative strengths, and the 
amplitude of the spectral break across the Balmer limit, are sensitive
to the mix of stellar ages.  The broadband spectral shape from these data is
less well determined, relying on secondary information from a KPNO
observation with a different slit area and with the accuracy of correction
for atmospheric dispersion an issue as well. We take the overall continuum
shape, integrated over the system, from large-aperture photometric
measures, as fit by {\it hyperz} \citep{bol00}. As discussed earlier, the best 
fit to the broad-band magnitudes for a single population from the
{\it hyperz} library is for continuous star formation. In comparison, the 
best single-age fit to the
GMOS spectral data, using the Starburst99 solar-abundance
spectra \citep{l99} ranges from 100-200 Myr depending
on how we weight various pieces of the spectrum. The predicted broadband
spectral distributions for these two cases are quite similar, as
one would expect from the strong weighting of the continuous star-formation
case to its youngest constituents.

\begin{figure}
\plotone{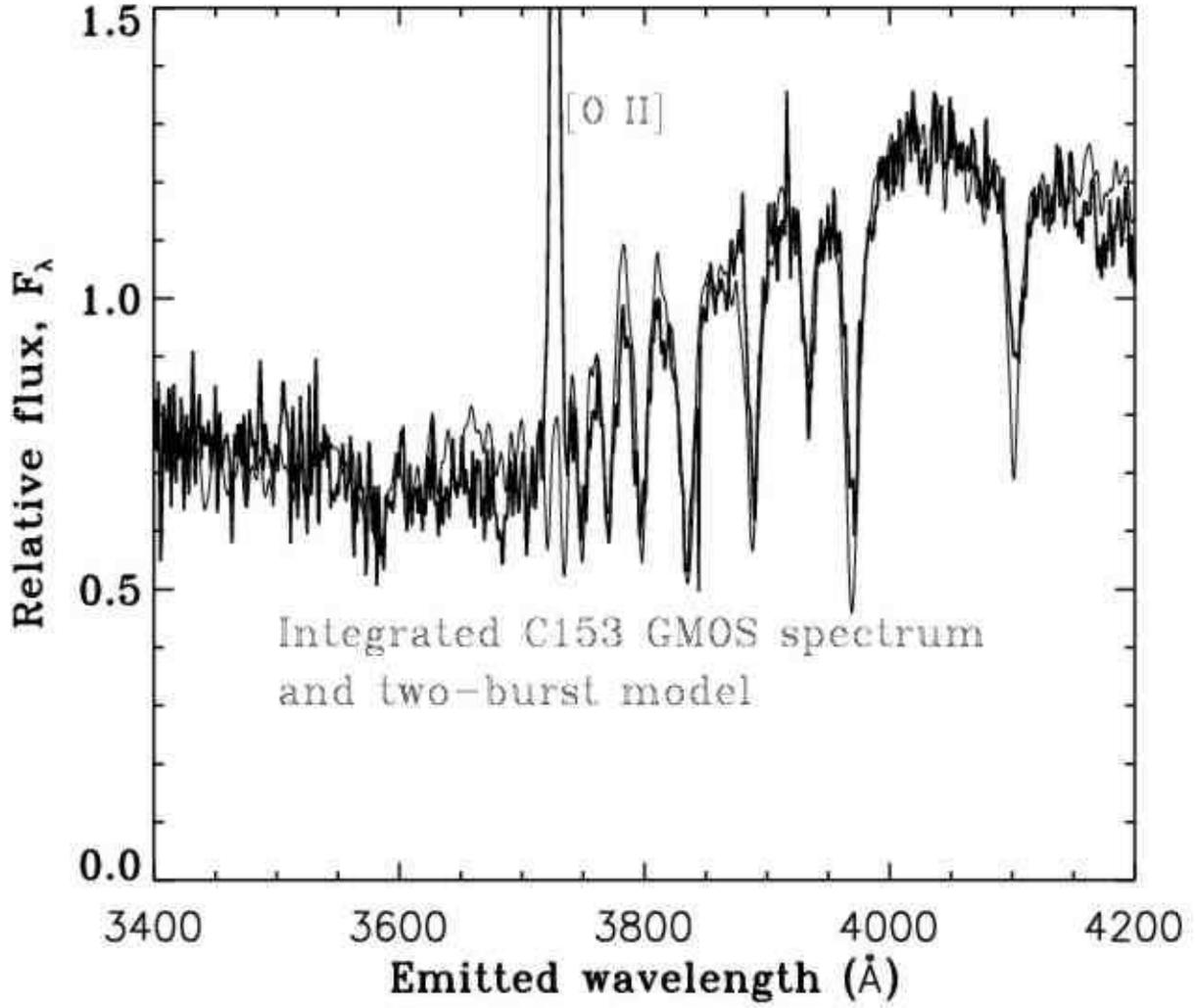}
\caption{GMOS blue spectrum of C153 with model spectrum
overplotted. \label{c153spec}}
\end{figure}

However for the GMOS spectrum, the continuum level between Balmer lines 
requires a multicomponent
population, as all the single-age burst populations and models with
continuous star formation give continuum levels here which are
too high to match the data by typically 15-20\%. In particular,
the continuum level just blue-ward of the H$\epsilon$+Ca II blend
suggests a very diluted 4000-\AA\  break, such as would be produced
by a small fraction of the light coming from an older population.
To narrow the range of parameters in the two-population fit, we took the old
population to have an age of 10 Gyr (based on the age of the Universe at the
epoch we observe Abell 2125 - from the WMAP cosmology, this would be 10.5 Gyr).
The best fit of such an old population and a younger starburst,
over the emitted wavelength range 3500--4200 \AA\  ,
has a burst of age 100 Myr contributing 90\% of the flux just longward
of the 4000-\AA\  break (specifically as measured at 4050 \AA\ ).
Over such a short wavelength span, its reddening is poorly determined,
being better constrained by the broadband continuum shape. 
In figure~\ref{c153spec} we show this fit overlaid on the observed
spectrum. This
model fits the broadband data as well as the single model from
{\it hyperz}, with the added feature of fitting the spectral data much
more closely. The longest two Balmer lines in this region
(H$\delta$ and H$\epsilon$) show some filling by emission, which is 
progressively
less important for the higher-order lines (since the emission decrement is
steeper than the absorption decrement). The best match between model
and data gives an absorption-line redshift $z=0.25275 \pm 0.00003$,
which will be the flux-averaged value along the slit.

In assessing the range of stellar populations that fit the GMOS data,
an ordinary $\chi ^2$ analysis is difficult because diagnostic
spectral features are concentrated in only a few wavelength ranges,
and we must allow for emission components to the Balmer lines and
interstellar absorption in, for example Ca II. We have used several
kinds of constraint to estimate error ranges. The wings of the Balmer
lines, beyond the line widths implied by other emission features, are
under-predicted by combinations with less than 0.6--0.9 of the 4050 \AA\ 
flux arising in a young population (for ages of that population from
100-25 Myr, respectively). A similar but weaker constraint comes from
not predicting Ca II $\lambda 3933$ stronger than observed; this is another
one-sided constraint, since interstellar absorption could strengthen
the line independent of the stars' properties. Fitting across the 
3400--4200 \AA\ emitted range, where most of the diagnostic absorption lines 
are and our flux calibration is most secure, we find that the young
population can be no older than 150 Myr or the observed spectrum becomes 
too blue. Between the Balmer lines H7-10, the continuum level is
best reproduced with a modest contribution ($0.3 \pm 0.1$ at 4050 \AA )
from an old population; the constraint is roughly $<0.8$ for all the
young ages we considered. Putting these together, the young population
implied by this model has an age $100 \pm 50$ Myr and contributes $0.7
\pm 0.1$ of the light emitted at 4050 \AA .

There is degeneracy between the age and reddening of the young population
(even to the extent that a such a simple two-component model is appropriate).
However, the continuum slope from 3300--3600 \AA  (short-ward of the
Balmer jump) is so blue that even the youngest burst populations
can be reddened by no more than E$_{B-V} = 0.1$, assuming a 
\citet{c00} effective extinction, without being redder
than we observe. This level of extinction applies strictly only to
the escaping light; as \citet{w92} stressed, the stars which
are most heavily obscured contribute least to the emerging colors. 
In view of the prominent dust structure seen in the WFPC2 images 
(e.g. figure~\ref{V47COL}),
we are certainly seeing only those stars which are behind the smallest
levels of extinction.

This model suggests that C153 is brighter now by a factor $\sim 10$ 
at 4050 \AA\  than it was before the onset of this star-formation
episode. In our model this episode has largely subsided, since the strong
Balmer absorption lines require a dropping star-formation rate
over much of the last $10^8$ years. Of course, given the apparent
large extinctions in the core of the galaxy, the single slit position
we have analyzed, and the relative
simplicity of our model, the most important point is that we see
evidence of star-formation activity on the time-scale of $10^8$ years.
This $10^8$ year scale is interesting, being comparable to the crossing time we
infer for this galaxy across the core ICM.

\subsubsection{Emission Line Ratios}

	To examine the energy sources powering the emission-line gas
which is prominent in our optical spectra, we have measured emission-line
ratios point-by-point along the slit for each spectrum. For the spectral region covered
in the higher-quality blue exposure, we have subtracted a stellar population
spectrum based on the best fit two-component model above, to minimize
the effects of underlying features from the starlight. Most line ratios
show radial gradients; given the seeing-limited resolution of these data,
such gradients might result from mixing of light from a core and surrounding
regions, as well as reflecting a more continuous change in conditions.
In the classification diagrams of
\citet{vo}, most of the regions we see 
fall among the LINERS but near the 
higher-ionization edge and toward the ``transition'' region which
occurs when AGN and starburst spectra are mixed. The ratios change as
one moves outward from the core in the sense that [O II]/[O III],
[S II]/H$\alpha$, and [O I]/H$\alpha$
increase while [O II]/H$\beta$ and [N II]/H$\alpha$ decrease. 
With [N II]/H$\alpha$=1.1, [O III]/H$\beta$=4.4, and [O I]/H$\alpha$=0.13,
the core spectrum would typically be classified as a reddened Sy\,2,
since the FWHM linewidth (from H$\beta$, where multiple structure is
least important) is about 350 km s$^{-1}$. Some of the line ratios
approach values typical of star-forming regions away from the nucleus
(such as [N II]/H$\alpha$), but [O I] is too strong all across the galaxy
to be accounted for solely by ordinary H II regions. The Balmer decrement
H$\alpha$/H$\beta$ reaches 10 at the center, declining to nearly
its recombination value $\approx 3$ at the edge of detected emission.

The [Ne III] lines at $\lambda \lambda$ 3889, 3969 \AA\  are detected near 
the edge of the
galaxy, although blended with Balmer lines so their intensities are
not precisely known (but near 0.15 of [O II]). These lines may be
observed at strengths including the values we observe in both AGN and starburst
systems. However, we do not detect [Ne V] $\lambda$ 3426, at the core or 
elsewhere, which would be a clear
sign of AGN photoionization, to a limit about 0.05 of [O II]. We are left
with the implication that ordinary star formation is not the dominant
source of ionization for the gas in C153. There may well be
a reddened contribution from an AGN, and the strong [O I] and [S II]
far from the core could implicate shock ionization as well.

The multiple kinematic components seen in some lines (especially
[O III]), which is strong and unblended) mean that all emission
lines are likely to be blends of multiple zones along the line
of sight, in which we do not have sufficient signal-to-noise
ratio to constrain the contributions from various velocities. 
This makes us reluctant to interpret the line
ratios in any greater detail.

\subsubsection{Velocity Field}

\begin{figure}
\epsscale{.90}
\plotone  {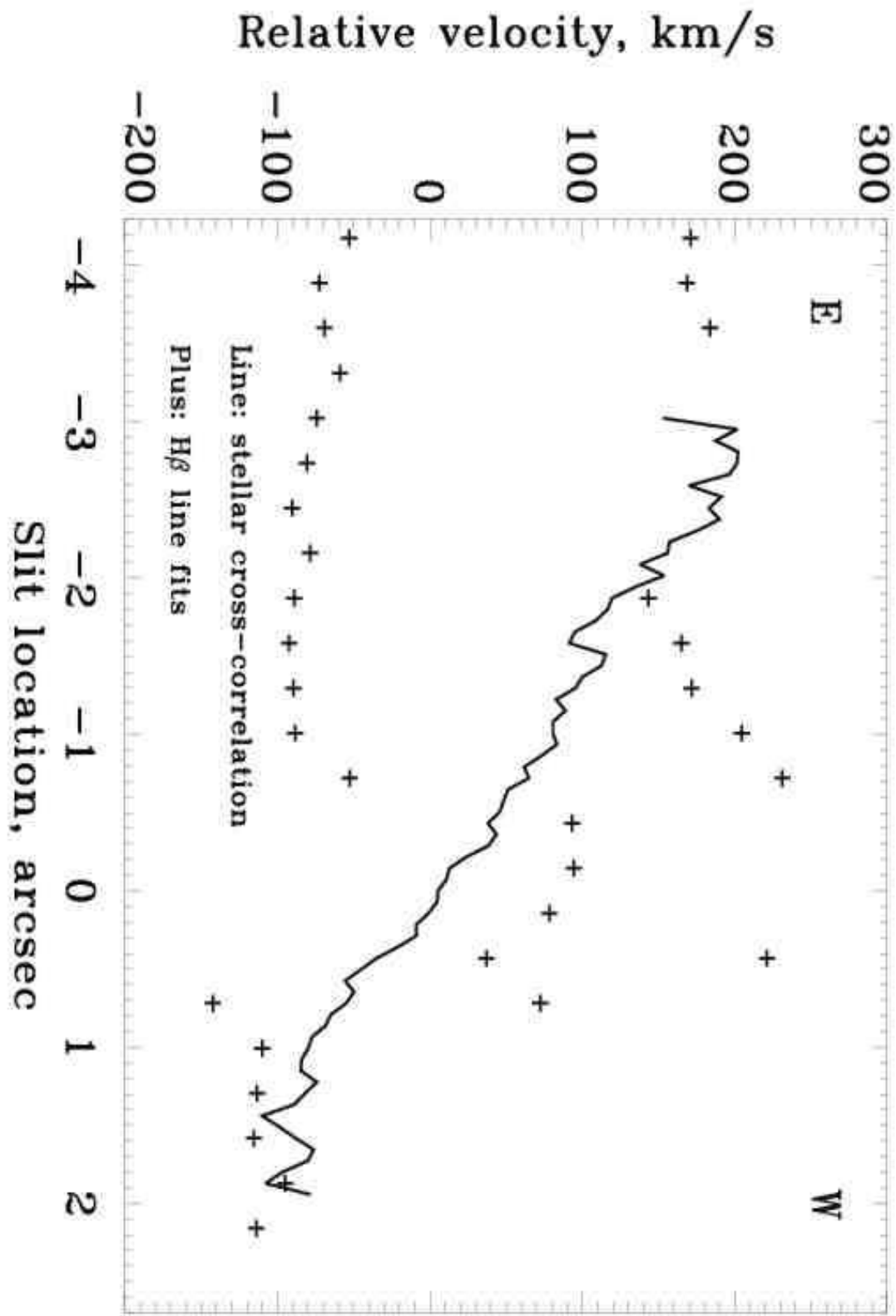}
\caption{Velocity behavior of stars and emission-line gas along the
slit of the blue GMOS spectrum. The solid line shows cross-correlation
velocities, pixel by pixel, where the error may be estimated from the
point-to-point scatter. Plus signs show H$\beta$ emission-centroids, after
binning by two pixels along the slit; adjacent points are
statistically independent. Multiple components were separated
by Gaussian deblending. \label{c153v}}
\end{figure}

We have evaluated the velocity structure of stars and emission-line gas along
the major axis of C153, as sampled by the GMOS spectra. For the stars,
we cross-correlated the peak spectrum with slices along the slit, using
the IRAF task {\it fxcor}. The result (figure~\ref{c153v}) shows a
nearly linear
position-velocity slice (noting that the seeing for the Gemini data
may have smeared away any small-scale rise in velocity near the core).
The full velocity range we see is 300 km s$^{-1}$ in the emitted
frame, with the
galaxy's eastern edge receding. The data span is truncated on the eastern end 
by the edge of the slit, rather than by signal-to-noise considerations.

The emission-line profiles are complex, with multiple components in
some locations. The cleanest results come from H$\beta$, since both line
blending and residual telluric absorption affect measurements of
H$\alpha$ and [N II], while [O III] is weaker. The results are shown
with the absorption-line data, in figure~\ref{c153v}.
The actual structure may be still more complex than the
two components we measure along some parts of the slit; the line profiles
are not completely fit by our pairs-of-Gaussians approach, but
anything more complicated risks becoming very strongly model-dependent.

We see regions in which the gas roughly follows the stellar velocities
(such as the area just E of the nucleus, and another one 1-2" W)
but additional components appear decoupled from the stars. Low gas velocities
persist from 1-4" east of the core, and very high velocities are 
seen $\approx 1$" east. These components may be loosely described
as counter-rotating, but this may be misleading since we do not know that
this gas is confined to the disk, or indeed still bound to the galaxy.
Such velocity structures could arise during stripping events either
as departing gas is decelerated to the ICM frame, a process that we view from
a poorly-constrained direction in this instance. Alternatively,
we could be seeing  some of the stripped gas falling back onto the 
galaxy, as occurs in the simulations by, for example,
\citet{v01} and \citet{s01}. The GMOS spectrum does
not extend far enough to the east (along the ``tail" of ionized and X-ray
gas) to show whether we do see gas reaching the local ICM frame at a
velocity offset of $\sim 1400$ km s$^{-1}$ from C153; the velocity range
projected against the disk is comparable to the maximum circular velocity. 

\subsection{C153 nucleus: star-formation or AGN ?}

In \citet{o99} we argued that C153 was an AGN from its optical
spectrum and the very large implied SFR from the radio. It also
seemed likely that the radio nucleus is quite compact, more consistent
with an AGN. For such a source one might expect the nuclear spectrum
to be flat or inverted, indicating a very small optically thick radio
AGN. However, from Gaussian fits using JMFIT in AIPS to our VLA 
D configuration radio images, we measure flux
densities of $6.88\pm 0.09$, $4.38\pm 0.03$ and $2.69\pm 0.22$ mJy
at 4860, 8460 and 14940 MHz. From Paper II, we found a 1400 MHz
total flux density for C153 of $22.94\pm .69$ mJy. These results
suggest a steep, optically thin radio core with a spectral index,
$\alpha$
of about 0.83, $S\propto \nu^{-\alpha}$ between 15 and 5 GHz. The
spectrum then steepens a bit to 0.96 between 5 and 1.4 GHz. At 1400 MHz about
95\% of the flux density was in a compact core while 5\% was extended
over about 14 arcseconds. While the combination of extended and
compact emission confuse slightly our knowledge of the spectrum
of the core of C153 alone, it is clear that the nuclear spectrum must
be a fairly normal optically thin synchrotron spectrum.

The {\it Chandra} X-ray observations do not detect the nucleus as
a point-like X-ray source. Using the 2 to 8 keV spectral range,
which is not confused by the lower temperature emission from the
diffuse tail, we can set a three sigma upper limit is 
$\sim 4\times 10^{41}$ ergs s$^{-1}$ for the 0.5 to 8 kev spectral
range. Thus there is no evidence for an AGN from X-rays. 

If we interpret the nucleus as star-formation we can use the radio
detection and X-ray limits to estimate the SFR, using the methods
described in Paper II, which result in equations 1 and 2 in that
paper.  These equations are based on the
calibration of the radio and X-ray luminosities for local 
star-forming galaxies in terms of a SFR. If star-formation were responsible 
for the
radio luminosity, then we would expect these two estimates of the SFR
to give similar results. The radio estimate gives a SFR of 
390 M$_\sun$ yr$^{-1}$ while the X-ray estimate gives $<30$ 
M$_\sun$ yr$^{-1}$. Although both relations allow for some
scatter about the fitted relations, these results are inconsistent with
each other and, along with the emission line strengths discussed by
\citep{o99}, suggest that star-formation is unlikely to be the origin of 
most of the radio emission. 

Given the GMOS results discussed above, the nucleus of C153 seems
most consistent with a heavily obscured, X-ray weak AGN. Certainly,
in this case it is possible that we are looking at a very Compton
thick environment around the nucleus of C153 which is hiding the
X-ray AGN emission. The very large HI optical depth seen against the
nucleus of  $\tau \sim
0.35$ \citep{dow} is certainly
consistent with such a picture. However, the radio nucleus is
optically thin, suggesting
that higher resolution observation might well reveal a resolved jet.

\section{Discussion}

\subsection{Violent Galaxy Stripping during a Cluster Merger}

	From the radio and optical data, the galaxy C153 stands
out as an unusual system. The galaxy appears to have undergone
a major disturbance. From the radio perspective, the galaxy
appears to be an AGN based on the compact core and its radio
luminosity. However, the optical spectrum show both indications
of weak AGN and star-formation activity. The alignment of some of the
fainter diffuse radio emission with the blue light from the
galaxy is also consistent with star-formation activity. The SED
fitting is consistent with star formation activity of the last
$10^8$ years. 
Clearly some very unusual event is taking place in this galaxy.

	The alignment of the emission from C153 is also suggestive
of an interaction with the cluster core. The elongated tail of
optical, [O II] and radio emission appears to point back to the cluster
core. Of course, C153 could just be an outlying member of A2125
seen in projection and the alignment could be chance. In this case,
C153 might be
a galaxy in the outskirts of the cluster which has undergone
a major galaxy-galaxy merger. However, the close proximity of
C153 to the central triple, along with the apparent alignment
of the system with the cluster core and the tail of bright
X-ray and optical emission suggests that the galaxy is probably
near the cluster core. Two mechanisms seem likely to account for
the peculiar galaxy. First a close passage to the massive cluster
core would tidally heat and disrupt a smaller galaxy \citep{hb,gn03a}.
Second, such a close passage through a moderately dense cluster
IGM would strip much of the ISM from the unfortunate galaxy
\citep[e.g.][]{a03}.

	Another piece of the puzzle is the relative velocities of
C153 and the central triple system. 
This pattern, unless it is a very unlikely projection effect, suggests
a complex three or four body interaction. \citet{m03} show that
the dynamics of the A2125 are consistent with a major cluster-cluster
merger seen in our line-of-sight which is near core passage. The radial
velocity of both 00057 and C153 are consistent with the higher
velocity component of the cluster-cluster merger, while 00105 and
00106 appear to be part of the lower velocity component. Thus it
seems possible that the activity in the core of A2125 is due to the
merger of two major subclumps in the A2125 cluster-cluster merger.

	On the somewhat large scale shown in figures~\ref{H1}, \ref{H2} and
\ref{X}, the X-ray and radio galaxy, 00039, with the large optical
halo to the west of the cluster core has a radial velocity consistent
with the main cluster system and the 00105/00106 pair. However, the
peak of the central diffuse X-ray emission is centered on 00057. 
Furthermore, the diffuse light is not centered of the central triple
but is offset toward 00057, suggesting a system not yet in equilibrium.
 Thus it would be very interesting to know
the velocity field of the central diffuse light. The overall pattern
we see in this cluster suggests a cluster-cluster merger taking place
along a major filament. Besides this large scale pattern, the
interaction seems to be manifesting itself as a number of smaller scale
events in the apparent core of the cluster. 

	C153 itself is very unusual. It has the brightest compact
radio source in the cluster. Its radio luminosity seems much too
high to be due to star-formation. The line ratios from the 4m
spectrum are most consistent with a weak AGN \citep{o99}.  However, its 
almost total lack of a 4000\AA\ break, its optical/NIR SED, and the
optical images suggest active star-formation. The ``flying fish''
appearance of the HST images suggest that something special has
happened to C153. The X-ray emitting tail also is far from the norm.
This feature is hard to understand if the galaxy is not moving
at high velocity through a relatively dense external medium.
Thus it seems likely that we are seeing C153 at some special point
in its evolution. There are many processes which might contribute
to the phenomena we observe. 

	Given the peculiar radial velocity field of C153 and the
central triple, it seems quite possible that some complex interaction
has occurred recently among these four galaxies. If the matter
responsible for the dominant potential near the cluster core is near
that of the dominant component seen by \citet{m03}, then C153 has
a large radial velocity relative to this system 
(at least 1400 km/s, perhaps much greater depending on its transverse 
motion). 
 However, the X-ray
tail suggests significant velocity in the plane of the sky as
well. Thus the total velocity of C153 appears to be large with respect
to the sound speed, $\sim 500$ km s$^{-1}$ \citep{w03}, perhaps Mach
4.   

If the cluster core has a
potential like most lensing systems, C153 will have experienced
significant tidal acceleration during its core passage \citep{hb}.
If the linear blue features seen near the cluster are gravitational
arcs, then we have direct evidence of such a potential.
If C153 plunged quite close to the central mass concentration of the
main component of Abell 2125, this encounter would have driven a strong tidal impulse 
on its disk. Because its velocity relative to the cluster
is much higher than its rotational velocity, such a tidal effect would
be impulsive, producing a stretching force within the disk \citep{hb}; 
a strongly peaked potential may in fact drive up the rate of
star formation through lateral compression of the disk and accompanying
changes in the rate of cloud collisions within the disk. Interstellar 
matter pulled outward by tidal effects would also become more vulnerable to
ram-pressure removal. If C153 has a velocity in the plane of
the sky of 1400 km s$^{-1}$ and has passed near the cluster core
then $\sim 10^8$ years may have passed since core passage.  
Thus C153 has had plenty of time to experience increased mass loss
from evolving stars into its ISM due to increased star formation. 

	Stripping of C153's ISM is the most obvious origin of
the X-ray tail of C153. Several authors have numerically modeled the 
stripping expected for the case of a spherical galaxy falling into
a cluster numerically (e.g. \citet{g87,b94,s99,a03}).
Given C153's relatively modest mass based on its 
rotation curve, its optical luminosity below $L_{*}$
and its linear size, the X-ray luminosity of $5\times 10^{41}$ erg s$^{-1}$
is rather high compared with the simulation of \citet{a03}. However,
the velocity of C153 relative to the external medium appears higher
than was modeled. Also the size of the unstripped region
may be larger due to the higher SFR and thus the internal mass loss
rate. C153 
appears to have a much more elongated potential based on its optical
isophotes but in our ground-based K-band imaging, the galaxy
appears rounder; thus the galaxy's 
cross-section is uncertain.  Higher resolution NIR imaging would
help us understand the galaxy's underlying potential. Also C153's current
trajectory may have originated from a much more violent event than
has been assumed in the models. Thus how the details of the C153
encounter relate to the models is uncertain. However, we can use them
as a guide to get some idea of what might be expected and how C153
needs to be different to fit the physical picture. 

	As can be seen in the simulation of \citet{b94}, the velocity
of material as it is stripped from the galaxy is much smaller
than that of the galaxy through the IGM. Furthermore, the flow is
unstable both as it originates from the galaxy and at the boundary of
the flow where Kelvin-Helmholtz instabilities are important. In these
regions conditions may be right for the formation of stars.

At the  ``leading edge" of the disk, the GMOS spectrum suggests that
star formation has shut down at
the observed epoch since these regions lack emission-line gas. 
Gas removal by stripping of C153 may be responsible. The velocity structure 
we see in the
emission-line gas shows enough complexity to fit either with the
episodic stripping from disks seen in the models by \citet{v01},
or by some of the stripped gas falling back into the disk
(as seen in the same models). We can make only very crude estimates of
the mass we see in gas in the stripped tail or plume. For very rough
guesses of the density ($n_e=10$ cm$^{-3}$)
in the [O II] region (and for solar oxygen
abundance), the mass of gas in this structure is about $10^8$ solar masses.
Still lower densities give larger masses. The mass in this volume of
the X-ray tail is likely to be several times larger.

In contrast to much of the galaxy disk itself, we do see evidence for
recent or ongoing star formation in the stripped material. Luminous
blue knots are seen on the ``downstream" side, and the rest-frame, near-UV
surface brightness of the plume (e.g. figures~\ref{V47},\ref{V47COL}) is 
difficult to explain unless
it contains stars so young that they must have been formed after the
gas left the galaxy. Stripped gas will be subjected to multiple and competing
physical effects whose interplay has yet to be properly understood (and
for which C153 should prove to be a valuable test case). In particular,
the observations of both optical line emission and X-rays from the
stripped region may make it possible to address the thermal
history of stripped gas. There could be a role for the ablation of dense
clouds by thermal conduction from the ICM, while the density
increase would make an interface cool the ICM efficiently; magnetic
effects could further complicate the behavior by introducing positional
and directional dependences in how efficient these processes are.

The event we have described suggests such a close passage to the 
nucleus of a rich cluster could be one of the ways cluster galaxies
change their morphology. It seems likely that C153 will eventually
settle down to look more like an ordinary galaxy, perhaps an
elliptical. Given the short timescale ($\sim 10^8$ years) for the event, 
we would only rarely expect to catch such an event in the act. As we
argued in Paper II, a cluster-cluster merger near core passage may be
a special time for galaxy evolution and C153 may be another way
that accelerated evolution can occur.

The unusually high velocity of C153's plunge through the densest part of
Abell 2125 lets us see the results of ram pressure in an unusually pure
way, and offers the possibility that further kinematics observations
can exploit the implicit mapping between location and time in the
stripped tail to further probe the history of gas after leaving the
galaxy. This system embodies in an extreme form processes that are
likely to have influenced the evolution of many cluster members in more
subtle ways.

\subsection{A Toy Model}

We can consider two pieces of the physical picture we have for
C153 in more detail: 1) the tidal effects of the close passage to
to the cluster center and 2) the stripping process.

\subsubsection{Assumptions}

The outermost part of the GMOS velocity slice are consistent with
a mass for C153 within 12 kpc of $9 \times 10^9$ M$_{\sun}$, assuming
a thin rotating disk. We assume that the blue features in figure
10 are gravitational arcs; then we can derive the mass within 65 kpc of the 
cluster
core in projection is $\sim 4.5 \times 10^{13}$ M$_{\sun}$ 
\citep{nb}. Conveniently, these values are consistent with the
parameters assumed by \citet{hb} for potential a) as well as the
mass of the galaxy they consider. Given that C153 has a relatively
long tail as seem in the plane of the sky and a large velocity offset
from the cluster mean, we assume that the galaxy is moving at an
angle to the line-of-site of 45\arcdeg, giving a velocity to the
cluster of $\sim 2000$ km sec$^{-1}$. This velocity results in
a time to cross a $\sim 250$ kpc cluster core radius
\citep{w03} of $\sim 1.3 \times 10^8$ years.

\subsubsection{Tidal Effects}

As observed C153 is about 100 kpc from the cluster core in projection.
If this distance is approximately the distance of the galaxy's closest
approach to the cluster core, then figures 2 and 3 of \citet{hb} shows
that at this point in its orbit the transverse and radial tidal acceleration
dominates over C153's internal forces by a factor $\sim
3$ and $\sim 10$ respectively. Thus the tidal acceleration dominates
the internal forces and the galaxy should be strongly disturbed and we
should expect the SFR rate to be stimulated. However, the timescale of
these large forces with respect to a galactic rotation of C153 is
short, so the effect of the these forces should be relatively local in
the galaxy.

Given the timescales involved, we would expect increased mass loss
from the evolving stellar populations to replenish the galaxy's ISM
and provide a continuous supply of gas to be stripped. The spectral 
modeling suggests that the bulk of the star-formation activity 
happended within the last $10^8$ years perhaps while C153 was at an
optimal point for the tidal effects to have their maximal affect.

\subsubsection{Stripping}

We can use the simple formalism from \citet{jo} to evaluate what might
be implied about the galaxy from the stripping we observe under our
assumptions and assuming simple spherical geometry for C153's ISM.
Instead of mass-loss rates from old stellar populations, we generalize
 their results using the results summarized by \citet{v05} where
the mass-loss rates are taken from the STARBURST99 models \citep{l99}.
Using equation 4 from \citet{jo} and assuming mass-loss-rates we
estimate the stagnation radius, roughly the radius of the tail just
behind the galaxy, to be 
$$R_{kpc}=8.7{\Bigl( \frac{S v_{\star 3}}{n_{-3}(v_{g3})^2}\Bigr) }^{1/2} $$
where the S is the SFR measured in solar masses per year, $v_{\star 3}$ is the
galaxy's effective velocity dispersion in 1000 km s$^{-1}$ ,
$n_{-3}$ is the IGM density in units of $10^{-3}$ particles cm$^{-3}$, and
$v_{g3}$  is the galaxy's velocity relative to the
IGM in units of 1000 km s$^{-1}$. We can roughly estimate $R_{kpc}$ from the
HST image and the group-based [OII] and U-band images to be about
12kpc. With $v_{\star 3} \sim 0.25$,$n_{-3}=1.7$ at 100kpc from the
cluster core \citep{w03}, and $v_{g3} \sim 2$, we get that the needed
SFR would be about 50 $M{\sun}$ yr$^{-1}$. However, the pressure at the ISM/IGM
boundary could also be affected by the excess energy in the mass-loss
outflows and/or by any AGN mechanical energy. Thus a lower SFR is
certainly possible, especially if the AGN is an important part of the
process. Also from \citet{jo}, the e-folding time for
our model galaxy to lose its ISM is $\sim 6 \times 10^{7}$ years, so
the mass-loss due to stripping would be dominated by the mass-loss
from the stellar population created during the tidal encounter.

This model is very simple, a ''toy'' model and there are several
uncertainties which
could change the actual parameters. However, it supports the
plausibility of the interpretation we have presented.

\section{Conclusions}

We have used optical, X-ray, and radio properties of galaxies near the
core of the populous and apparently merging cluster Abell 2125 to
probe the effects of cluster-scale events on individual galaxies. The
radio source C153 shows spectacular features which suggest both
an AGN and a starburst have been triggered during the interaction.
A tail of emission, seen in X-rays, [OII] line emission and
UV-light are also seen. A ``toy'' model suggests the event is
consistent with triggering
by tidal effects, and stripping of gas during the
last 10$^8$ years, roughly the time it would take to cross the
cluster core at an unusually high relative velocity.

Individual events of this intensity may be rare, requiring a gas-rich
galaxy to cross deep within the cluster core at very high velocity.
However, they do furnish an opportunity to test our ideas about
how stripping proceeds, with less confusion from purely gravitational
effects than is generally seen in more protracted instances. In
particular, simultaneous detection of optical and X-ray gas in the
plume from C153 in A2125 may show us the thermal history of gas as
it moves from the interstellar medium to the hot intracluster medium.
This event may also be showing us another way in which galaxy
evolution can occur.

\acknowledgments
This research was supported by NASA through STSCI grant GO-07279.01-96A.
We thank David Bohlender for his assistance in verifying the observational
parameters for the GMOS spectra. We also thank Jean Eilek for comments
on the text.

\clearpage

\begin{deluxetable}{lllllllll}
\tablecolumns{9}
\tablewidth{0pt}
\tablecaption{A2125 Central Galaxies Properties\label{RP}}
\tablehead{
\colhead{Name} &
\colhead{Coord Name} & 
\colhead{Size} &
\colhead{$\log(L_{20cm}$)} &
\colhead{$M_R$} &
\colhead{Spec T.} &
\colhead{Age} &
\colhead{$A_V$} &
\colhead{z}} 
\startdata
00047&154109+661544&53.6&24.59&-21.7&8&3.5&1.0&0.2528e\\
00057&154114+661557&10.7&23.69&-23.3&1&5.5&0.4&0.2518 \\
00105&154114+661603&15.3&23.06&-23.1&1&7.5&0.4&0.2470 \\
00106&154115+661556&88.1&24.42&-22.7&1&9.5&0.0&0.2466 \\
\enddata
\end{deluxetable}

\clearpage

\clearpage

\end{document}